\PassOptionsToPackage{numbers,square,sort&compress}{natbib}
\documentclass{iopjournal}
\usepackage{graphicx}%
\usepackage{multirow}%
\usepackage{amsmath,amssymb,amsfonts}%
\usepackage{amsthm}%
\usepackage{mathrsfs}%
\usepackage[title]{appendix}%
\usepackage{xcolor}%
\usepackage{textcomp}%
\usepackage{manyfoot}%
\usepackage{booktabs}%
\usepackage{algorithm}%
\usepackage{algorithmicx}%
\usepackage{algpseudocode}%
\usepackage{listings}%
\usepackage{verbatim}
\usepackage{caption}
\usepackage{subcaption}
\usepackage{float}
\usepackage{siunitx}
\usepackage[dvipsnames]{xcolor}

\usepackage{tikz}
\usepackage{url}
\usepackage{pgfplots}
\usepackage{lmodern}

\usetikzlibrary{calc, positioning, shapes.geometric, decorations.markings, arrows.meta}
\usepgfplotslibrary{groupplots}
\usepgfplotslibrary{fillbetween}

\pgfplotsset{
  compat=1.18,
  width=7cm,
  height=4.5cm,
  every axis/.append style={
    grid=both,
    grid style={gray!20},
    minor tick num=1,
  },
  legend style={
    font=\small,
    cells={anchor=west},
    /tikz/every even column/.append style={column sep=0.4em},
  },
}
\tikzset{
    neuron/.style={
        circle,
        draw=black!75,
        thick,
        minimum size=5.5mm,
        inner sep=0pt,
        font=\small\sffamily,
        align=center,
        fill opacity=0.3
    },
    link/.style={
        -{Stealth[length=2mm,width=2mm]},
        thick,
        draw=black!75
    }
}

\definecolor{node0color}{RGB}{128, 0, 32}    
\definecolor{node1color}{RGB}{211, 47, 47}   
\definecolor{node2color}{RGB}{245, 124, 0}   
\definecolor{node3color}{RGB}{195, 130, 0}   

\newcommand{\netN}{N}                             
\newcommand{\netT}{T_{\mathrm{tot}}}              
\newcommand{\netPhaseProf}{\Phi}                  
\newcommand{\netPhase}{\varphi}                   

\newcommand{\nodeDelay}[1]{\Delta_{#1}}            
\newcommand{\node}[1]{n_{#1}}                      
\newcommand{\nodeDelayRel}[1]{\netPhase_{#1}}          
\newcommand{\delayInf}{\Delta_{\infty}}           
\newcommand{\inputPeriod}{T_{\mathrm{in}}}        
\newcommand{\restPeriod}{T_{\mathrm{r}}}          
\newcommand{\minPeriod}{T_{\mathrm{min}}}         

\newcommand{\eDF}{\mathrm{eDF}}                    
\newcommand{\ePRC}{\mathrm{ePRC}}                  
\newcommand{\nominalT}{T_{\mathrm{nom}}}       
\newcommand{\perturbT}{T_{\mathrm{p}}}        
\newcommand{\phaseOffset}{\delta_{\mathrm{p}}}        
\newcommand{\phaseShift}{\Pi(\inputPeriod,\phaseOffset)}        
\newcommand{\delayFunc}{\Delta(\inputPeriod)}     
\newcommand{\phaseFunc}{\varphi(\inputPeriod)}    
\newcommand{\eventIn}{\mathcal{E}^{\mathrm{in}}}       
\newcommand{\eventP}{\mathcal{E}^{\mathrm{p}}}         
\newcommand{\eventOut}{\mathcal{E}^{\mathrm{out}}}     
\newcommand{\timeIn}[1]{t^{\mathrm{in}}_{#1}}     
\newcommand{\timeP}[1]{t^{\mathrm{p}}_{#1}}     
\newcommand{\timeOut}[1]{t^{\mathrm{out}}_{#1}}   
\newcommand{\diffT}{\perturbT - \nominalT}

\newcommand{\gNa}{g_{\mathrm{Na}}}               
\newcommand{\gK}{g_{\mathrm{K}}}                 
\newcommand{\gA}{g_{\mathrm{A}}}                 
\newcommand{\gSyn}{g_{\mathrm{syn}}}             
\newcommand{\tauDecay}{\tau_{\mathrm{syn}}}    
\newcommand{\iA}{I_{\mathrm{A}}}              
\newcommand{\iNa}{I_{\mathrm{Na}}}              
\newcommand{\iNaReb}{I_{\mathrm{Na,Reb}}}              
\newcommand{\iCaT}{I_{\mathrm{Ca,T}}}              
\newcommand{\iSyn}{I_{\mathrm{syn}}}          
\newcommand{\vTh}{V_{\mathrm{th}}}           

\begin{document}

\articletype{Paper} 

\title{Neuromorphic design of a rhythmic network}

\author{Robin Wroblowski$^{1,*}$\orcid{0009-0008-6764-2813} \;\;\; and Rodolphe Sepulchre$^{1,2}$\orcid{0000-0002-7047-3124}}

\affil{$^1${KU Leuven}, {Departement of Electrical Engineering (ESAT)}, {STADIUS Center for Dynamical Systems, Signal Processing and Data Analytics}, Leuven, Belgium}

\affil{$^2${University of Cambridge}, {Department of Engineering}, Cambridge, United Kingdom}

\affil{$^*$Author to whom any correspondence should be addressed.}

\email{robin.wroblowski@kuleuven.be, rodolphe.sepulchre@kuleuven.be}


\begin{abstract}
    Central Pattern Generators (CPGs) have long inspired the design of rhythms that combine endogenous generation and exogenous interaction.
    In neurodynamics and engineering studies, CPGs have been primarily analysed and designed as networks of  coupled oscillators.
    In contrast, the present article regards CPGs as networks that propagate events through excitable rather than oscillatory nodes.
    The proposed event-based perspective reduces the design of network-level rhythmic properties to the tuning of local nodal input--output characteristics. 
    This reduction enables a distributed and robust design methodology: nodes can be tuned independently, they can be heterogeneous, and the global rhythm results from local event propagation rather than from a specific global tuning of the network parameters.
    The present article builds on the definitions introduced in our preliminary work~\cite{wroblowskiDescribingFunctionsPhase2025} that parallel existing characterizations of oscillators: the event describing function ($\eDF$), which characterises the propagation delay of a node, and the event phase response curve ($\ePRC$), which characterises the phase sensitivity of a node.
    Event-based CPGs made of excitable rather than oscillatory nodes emerge as a principled design framework for robust rhythm generation.
\end{abstract}

\section{Introduction}\label{sec:intro} 

    Rhythmic motor behaviours, including locomotion, respiration, and swallowing, rely on the repeated and coordinated activation of muscles. 
    Biological systems typically govern these movements through central pattern generators (CPGs): neural circuits capable of producing rhythmic motor commands without requiring continuous rhythmic input~\cite{brownNatureFundamentalActivity1914, marderPrinciplesRhythmicMotor1996a}. 
    While sensory feedback and descending signals modulate the resulting behaviour, CPGs provide the fundamental organizing principle for motor control~\cite{selverstonOscillationsOscillatoryBehavior2006, calabreseDegenerateNeuronalCircuit2025}. 
    Classical interpretations of biological CPGs distinguish between rhythms driven by intrinsic pacemakers and those emerging from network interactions~\cite{grillnerBiologicalPatternGeneration2006, marderCentralPatternGenerators2001}. 
    Pacemaker-driven architectures build upon intrinsically oscillatory neurons coordinated through synaptic coupling~\cite{ramirezPacemakerNeuronsNeuronal2004}. 
    Conversely, network-driven circuits do not require individual neurons to oscillate in isolation; the rhythm arises entirely from structural interactions like reciprocal inhibition and post-inhibitory rebound~\cite{marderCentralPatternGenerators2001}. 
    Actual biological networks likely blend these mechanisms, with individual neuronal contributions shifting based on operating points, synaptic inputs, and neuromodulatory states~\cite{ramirezPacemakerNeuronsNeuronal2004}.   Such architectures have deeply influenced robotic locomotion by providing a low-dimensional parametrisation for coordinated, high-dimensional movement~\cite{ijspeertCentralPatternGenerators2008a, yuSurveyCPGInspiredControl2014}.
    
    Traditionally, engineers have favoured the design of rhythmic actuator trajectories from the theory of oscillator networks~\cite{collinsCoupledNonlinearOscillators1993, collinsGrouptheoreticApproachRings1994}. 
    Translating the biological CPG concept into robotics predominantly yields networks of coupled limit-cycle oscillators, whether formulated through abstract models (e.g., Hopf, Van der Pol) or paired half-centers abstracted as effective oscillatory units (e.g., Matsuoka circuits)~\cite{ijspeertSimulationRoboticsStudies2005, yuGaitGenerationApproach2026}.
    In these architectures, the continuous nodal dynamics and coupling topology directly dictate synchronization patterns, phase relationships, and viable gait transitions.
    However, endogenous rhythm generation, while crucial, solves only one piece of the practical locomotion puzzle~\cite{kuoRelativeRolesFeedforward2002, ryuOptimalityPrincipleLocomotor2021, ijspeertIntegrationFeedforwardFeedback2023}. 
    Movement initiation, adaptation, and termination critically rely  on sensory feedback, reflexes, body mechanics, and descending commands~\cite{gouldingSensoryFeedbackDynamic2025, bidayeSixleggedWalkingInsects2018a}. 
    Consequently, robotic and neuromechanical designs either investigate purely reflex-driven architectures~\cite{geyerMuscleReflexModelThat2010, songNeuralCircuitryThat2015} or augment feedforward CPG signals with proprioceptive feedback loops~\cite{proctorReflexesPreflexesRole2010, dzeladiniContributionCentralPattern2014, codianniRolesAscendingSensory2020, dirussoInvestigatingRolesReflexes2023, pazzagliaBalancingCentralControl2025}, exteroceptive sensing~\cite{saputraCombiningReflexesExternal2022}, and local reactive reflex pathways~\cite{ferreiraCombiningCentralPattern2015, akkawutvanichAdaptiveParallelReflex2020, yuEnhancingAdaptabilityLocal2020, schmidtAdaptingHighlyDynamicCompliant2021}.
    Yet, the ideal method for integrating internally generated rhythms with incoming sensory data remains an open design challenge, particularly in asynchronous or event-based control schemes.

    The event-based approach of neuromorphic control provides an alternative angle of attack. 
    The present paper revisits CPG design from the perspective of interconnecting excitable rather than oscillatory nodes, focusing on tuning and robustness while bypassing the physiological details of biological neurons.
    A loop of nodes that propagate events forms a feedback system that does not separate the mechanism of endogenous rhythm generation from the interaction with the environment.
    Event-based design is hierarchical, and it includes \textit{automation} (i.e., rhythm generation) and \textit{regulation} (i.e. interaction with the environment) at any scale~\cite{huoWinnerTakesAllMechanismEvent2025, sepulchreRegulationCalibration2025, schmetterlingNeuromorphicControlPendulum2024, ribarNeuromorphicControlDesigning2021, szczecinskiPerspectiveNeuromorphicControl2023}. 
    Formulating rhythmic motor control through discrete events structurally unifies the question of endogenous rhythm generation and the question of exogenous interaction, allowing nodal interactions and external perturbations to share the exact same event-based representation.

    Excitable neurons are natural elementary units for such event-propagating networks~\cite{sepulchreSpikingControlSystems2022a}. 
    Unlike an oscillator, the isolated behaviour of an excitable neuron is a stable equilibrium, the resting state,  until an input drives the behaviour above threshold, thereby triggering a transient event such as a neuronal spike or burst~\cite{hodgkinQuantitativeDescriptionMembrane1952, izhikevichDynamicalSystemsNeuroscience2006b}. 
    While the same neuronal model can be excitable or oscillatory in distinct parameter regimes~\cite{marderNeuromodulationNeuronalCircuits2012, yoshinoSyntheticAnalysisPeriodically1999}, these two operating states enforce fundamentally different mechanisms of network coordination. 
    Oscillatory nodes possess intrinsic frequencies prone to incompatibility and conflict under parameter heterogeneity. 
    As a consequence, networks of oscillators  demand precise parameter balancing, forcing the joint tuning of nodal and network parameters to achieve stable phase-locking. 
    Excitable nodes, in contrast,  lack an autonomous rhythm. As a consequence, they introduce no competing intrinsic frequencies that require global reconciliation.
    Global rhythmic behaviour instead emerges purely from the composition of local discrete-event propagation, as each node independently processes incoming events into outgoing responses~\cite{sepulchreControlScalesPositive2019a}.
    This isolates the tuning process, replacing the precise balancing of high-dimensional coupled differential equations with strictly local, independent design rules.

    Post-inhibitory rebound provides a highly effective mechanism for organizing this sequential event propagation~\cite{perkelMotorPatternProduction1974, satterlieReciprocalInhibitionPostinhibitory1985, dethierPositiveFeedbackCellular2015}. 
    Because release from inhibition can actively trigger a spike or burst, inhibitory synapses do more than merely suppress activity; they structure alternation and drive sequential propagation.
    Specifically, directed ring topologies of cyclic inhibitory connections can sustain robust rhythmic patterns even when every constituent neuron remains entirely quiescent in isolation. 
    Network and population models demonstrate that sequences, travelling waves, and low-dimensional rotational dynamics---from motor gaits to respiratory rhythms~\cite{DelNegroRespiratoryRhythmEmergent2002}---readily materialise without intrinsic pacemakers~\cite{komiSpatialNetworkPrinciples2024, londonoalvarezAttractorbasedModelsSequences2025, lindenMovementGovernedRotational2022}. 
    While many of these existing models capture collective activity through continuous firing-rate dynamics, our approach preserves the discrete, spike-based nature of neuronal signalling to frame rhythmicity purely as event propagation between excitable nodes. 
    We mirror the concept of wave propagation in excitable media~\cite{rinzelPropagatingActivityPatterns1998}, but channel it through discrete inhibitory interactions and post-inhibitory rebound within an explicitly defined, discrete network topology.

    Traditionally, the analysis tools of rhythmic networks have relied on the theory of limit-cycle oscillators. 
    In control theory, describing function analysis and  harmonic-balance methods have been central to predict oscillations in feedback loops~\cite{khalilNonlinearSystems2002, varigondaDynamicsRelayRelaxation2001, astromOscillationsSystemsRelay1995, wuFeedbackControlOscillations2014, iwasakiSensoryFeedbackMechanism2006a, futakataEntrainmentNaturalOscillations2011}. 
    Entrainment and synchronization questions have been instead studied via the phase-response curve analysis~\cite{sacreSensitivityAnalysisOscillator2014a}.
    Phase reduction maps an autonomous limit-cycle oscillator onto a single phase variable, and phase response curves (PRCs) quantify how external perturbations advance or delay the underlying cycle~\cite{winfreeGeometryBiologicalTime2001a, schultheissPhaseResponseCurves2012, nakaoPhaseReductionApproach2016,efimovControllingPhaseOscillator2009}. 
    This approach has been extensively adapted for various oscillator models, including relaxation systems~\cite{izhikevichPhaseEquationsRelaxation2000, sacreSingularlyPerturbedPhase2016}, excitable media via isostables~\cite{wilsonExtendingPhaseReduction2015}, and functional network responses~\cite{izhikevichDynamicalSystemsNeuroscience2006b, cuiFunctionalPhaseResponse2009}. 
    By isolating phase interactions, researchers can tractably predict synchronization, phase-locked states, and overall stability in coupled oscillatory networks~\cite{oprisanExistenceStabilityCriteria2010, canavierPhaseResponseCharacteristics1997, canavierPulseCoupledOscillators2010, zhuDynamicalAnalysisMultistable2026, achuthanSynapticIntrinsicDeterminants2011, kuramotoCollectiveSynchronizationPulsecoupled1991}. 
    Yet, standard describing functions and phase response curves presuppose autonomous limit-cycle oscillations, making them inapplicable to excitable nodes that settle at a stable resting state.
    Sieling et al.~\cite{sielingPhaseResponseTheory2012} first demonstrated that phase response concepts could be extended beyond autonomous limit cycles to individual non-oscillatory neurons.
    Our recent work~\cite{wroblowskiDescribingFunctionsPhase2025} shifted this perspective from isolated cells to fully excitable, non-oscillatory networks by introducing the event describing function ($\eDF$) to quantify nodal propagation delay and the event phase response curve ($\ePRC$) to quantify phase sensitivity.
    The present article develops this $\eDF$ and $\ePRC$ foundation into a systematic design and regulation methodology for heterogeneous CPG architectures.

    The rest of the article is organised as follows:
    Section~\ref{sec:motivating_example} uses a clock analogy to contrast coupled-oscillator and excitable networks, showing how event-propagating architectures decouple local timing adjustments and accommodate parameter heterogeneity.    
    Section~\ref{sec:endogenous} addresses endogenous rhythm generation by developing the event describing function ($\eDF$) as a design tool. 
    We demonstrate how nodal parameter variations shape the $\eDF$, enabling the systematic design of heterogeneous networks with prescribed properties, such as the total network period and phase profile.    
    Section~\ref{sec:exogenous} studies exogenous rhythm regulation using the event phase response curve ($\ePRC$). 
    We show how nodal parameters shape the $\ePRC$ under excitatory and inhibitory perturbations, providing a tractable local methodology for designing sensory sensitivity and network entrainment.    
    Finally, Section~\ref{sec:conclusion} summarises the results and outlines directions for future work.


\section{Motivating example}\label{sec:motivating_example}   

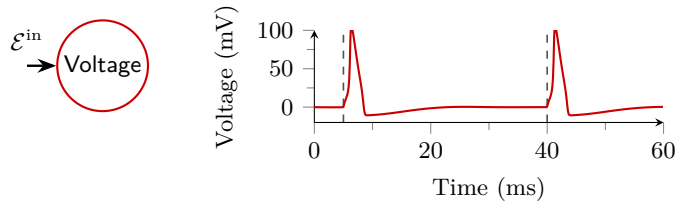
\begin{figure}[htbp]
\centering
\begin{tikzpicture}[
    font=\sffamily\small,
    arrow/.style={-Stealth, thick, line width=0.9pt},
    circle_node/.style={draw, circle, minimum size=1.2cm, align=center, inner sep=1pt, line width=0.8pt}
]


\node[font=\sffamily\bfseries\small, anchor=west] at (-2.5, 1.2) {(a) Autonomous Oscillator};

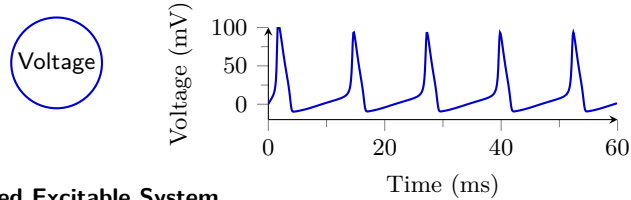
\node[circle_node, draw=blue!80!black, text=black] (osc_node) at (0, 0) {Voltage};
\begin{axis}[
    at={(2.8cm, -0.75cm)},
    width=6.2cm,
    height=2.8cm,
    axis lines=left,
    grid=none,
    xlabel={Time (ms)},
    ylabel={Voltage (mV)},
    xmin=0, xmax=60,
    ymin=-20, ymax=100,
    tick align=outside,
    tick label style={font=\small},
    label style={font=\small},
    enlargelimits=false
]
    \addplot[blue!80!black, thick, no markers, each nth point=5] 
        table [x=time_ms, y=v_oscillatory_mV, col sep=comma] {plotdata/neuron_simulation_data.csv};
\end{axis}


\node[font=\sffamily\bfseries\small, anchor=west] at (-2.5, -1.8) {(b) Triggered Excitable System};

\node[circle_node, draw=red!80!black, text=black] (exc_node) at (0, -3.0) {Voltage};

\draw[arrow] (-1, -3.0) -- (exc_node.west) 
    node[at start, above=2pt, font=\sffamily\small] {$\eventIn$};

\begin{axis}[
    at={(2.8cm, -3.75cm)},
    width=6.2cm,
    height=2.8cm,
    axis lines=left,
    grid=none,
    xlabel={Time (ms)},
    ylabel={Voltage (mV)},
    xmin=0, xmax=60,
    ymin=-20, ymax=100,
    tick align=outside,
    tick label style={font=\small},
    label style={font=\small},
    enlargelimits=false
]
    \draw[dashed, opacity=0.7, line width=0.6pt] (axis cs:5, -20) -- (axis cs:5, 100);
    \draw[dashed, opacity=0.7, line width=0.6pt] (axis cs:40, -20) -- (axis cs:40, 100);

    \addplot[red!80!black, thick, no markers, each nth point=5] 
        table [x=time_ms, y=v_excitable_mV, col sep=comma] {plotdata/neuron_simulation_data.csv};
\end{axis}

\end{tikzpicture}
\caption{Comparison of a Hodgkin-Huxley neuron used as: (a) a closed, autonomous oscillator (constant applied current) vs. (b) an open, event-driven excitable system (pulsatile inputs). Incoming event sequence ($\eventIn$) indicated by vertical dashed lines.}
\label{fig:closed_vs_open_neuron}
\end{figure}

    The core premise of this work lies in a shift from classical coupled-oscillator networks to excitable networks. 
    This structural distinction is illustrated via a Hodgkin--Huxley neuron in Figure~\ref{fig:closed_vs_open_neuron}: panel~(a) displays an autonomous oscillatory regime generating a continuous limit cycle, whereas panel~(b) shows an excitable regime that remains quiescent at rest until an incoming input event triggers an output event.
    Mathematically, the two distinct operating regimes of the same model only differ by a constant input current. 
    Yet, the input--output properties of the two systems differ considerably, and they impact both the analysis and the design of the neuron in a neural network. 
    In particular, the excitable regime is a sweet spot between a system that can generate events autonomously and a system that interacts with its environment. 
    
    To suggest the practical differences between oscillator-based and network-based CPGs, we start from the elementary design question of interconnecting neurons to produce a clock-like cyclic rhythm, which will serve as a guiding analogy for the remainder of the paper.
    We first demonstrate that decoupling local from global behaviour grants the excitable network superior robustness against heterogeneity. 
    Following this, we contrast both paradigms directly by illustrating how a localised synaptic modulation propagates globally across the oscillatory network, yet remains isolated to local dynamics in the excitable ring.

    \subsection{Network design of a clock: sequential ring activation}  
        

        In the absence of interaction with the environment, a clock is  a dynamical system with a stable limit cycle characterised by the sequential activation of nodes in a predetermined firing order. 
        In an excitable network, a unidirectional ring topology enables this objective with minimal connectivity (Figure~\ref{fig:ring}, left).
        When an active node fires, it excites its downstream neighbour, establishing an event-driven propagation motif along the ring. 
        In this framework, a nodal spike serves as a discrete trigger for the state transition of a discrete automaton.  
        The continuous duration of an active state is governed by the delay between successive spikes.
        We can regard this motif as an abstraction of many biological CPGs, where neurons burst one after the other, and where the duration of the burst physically controls the duration and delay of the successive nodal events. 
        We will return to the extension of the neuromorphic framework to bursting networks in the conclusion.

        The left panel of Figure~\ref{fig:ring} illustrates the directed ring topology. 
        We adopt a four-node network ($\netN=4$) throughout this paper to maintain clear and legible visualizations. 
        This architecture can be interpreted as clustering the conventional 12-hour clock positions into four functional groups, an aspect detailed in Section~\ref{sec:3:clustering}. 
        The design principles presented here extend directly to rings of any size $\netN$.
        
        \input{fig2}

        The right panel of Figure~\ref{fig:ring} shows the resulting voltage traces, highlighting the activity of node $\node{0}$. 
        As illustrated, these excitable nodes rest at a stable subthreshold equilibrium until an incoming inhibitory event hyperpolarises them, triggering a post-inhibitory rebound spike upon release (this mechanism is explained in detail in Section~\ref{sec:3:mechanism}. 
        In this work, we focus on this fundamental ring topology, as it captures the fundamental event-driven foundation onto which more complex layers can be added.

        Attempting to realise a ring oscillator with oscillatory nodes reveals fundamental limitations. 
        When coupled with purely excitatory synapses, autonomous oscillators tend to lock in phase, collapsing the desired sequence into global synchrony.
        Conversely, inhibitory coupling can support phase-lagged sequential firing, but this solution is rarely unique; inhibitory oscillator rings frequently exhibit complex multi-stability~\cite{collinsGrouptheoreticApproachRings1994}.
        Consequently, the emergence of the correct sequence becomes heavily reliant on specific initial conditions, leaving the desired rhythm vulnerable to parameter variations and transient perturbations.

        To eliminate competing attractors and enforce the sequential firing order as the unique stable limit cycle in an oscillatory network, additional structural constraints are required. 
        Specifically, one must combine global winner-take-all (WTA) inhibition---to ensure solitary activation---with a unidirectional excitatory ring---to enforce the firing sequence. 
        An analogue of this architecture is seen in rebound winner-take-all (RWTA) networks~\cite{huoWinnerTakesAllMechanismEvent2025}, which apply this dual-layer topology to rebound-excitable neurons. 
        This composite architecture is straightforward to implement with excitable neurons, and can yield robustness benefits relative to the ring topology.
        Yet, a similar oscillatory network remains inherently fragile, requiring global and precise tuning of intrinsic frequencies and synaptic gains to maintain coherent phase-locking~\cite{smealPhaseresponseCurvesSynchronized2010a}. 
        As we will discuss in Section~\ref{sec:2:robust}, this makes the design problem similarly vulnerable to parameter variations and perturbations.

    \subsection{Robustness to heterogeneity}
    \label{sec:2:robust}
        Robustness to parameter variation is essential for a straightforward design that avoids precise fine-tuning. 
        Beyond simplifying implementation, this robustness directly dictates the network's tunability: the system's behaviour must remain qualitatively stable under parameter variations to allow for reliable modulation and control. 
        With this in mind, let us first examine how the two architectures compare in terms of robustness to parameter heterogeneity. 
        We deliberately benchmark the minimal excitable network against the most robust oscillatory configuration capable of producing a stable clock rhythm.

\begin{figure}[htb]
  \centering

  \begin{tikzpicture}
    \begin{axis}[
      width=8cm,
      height=4.5cm,
      enlarge x limits=0.05,
      ymin=0, ymax=105,
      xmin=0,
      xlabel={Heterogeneity $h$ (\%)},
      ylabel={Valid (\%)},
      tick label style={font=\footnotesize},
      label style={font=\small},
      legend style={font=\footnotesize, nodes={inner sep=2pt}},
      legend pos= north east, 
      grid=both,
      grid style={gray!15},
      every axis plot/.append style={line width=1pt},
    ]

    \addplot [name path=upper_osc, draw=none, forget plot] 
      table[x=heterogeneity, y=ci_upper, col sep=comma] {plotdata/plotdata_oscillatory_robustness.csv};
    \addplot [name path=lower_osc, draw=none, forget plot] 
      table[x=heterogeneity, y=ci_lower, col sep=comma] {plotdata/plotdata_oscillatory_robustness.csv};
    \addplot [fill=blue!20, forget plot] fill between[of=upper_osc and lower_osc];
    
    \addplot [blue, mark=*, mark size=1.5pt] 
      table[x=heterogeneity, y=mean, col sep=comma] {plotdata/plotdata_oscillatory_robustness.csv};
    \addlegendentry{Oscillatory}

    \addplot [name path=upper_exc, draw=none, forget plot] 
      table[x=heterogeneity, y=ci_upper, col sep=comma] {plotdata/plotdata_excitable_robustness.csv};
    \addplot [name path=lower_exc, draw=none, forget plot] 
      table[x=heterogeneity, y=ci_lower, col sep=comma] {plotdata/plotdata_excitable_robustness.csv};
    \addplot [fill=red!20, forget plot] fill between[of=upper_exc and lower_exc];
    
    \addplot [red, mark=square*, mark size=1.5pt] 
      table[x=heterogeneity, y=mean, col sep=comma] {plotdata/plotdata_excitable_robustness.csv};
    \addlegendentry{Excitable}

    \end{axis}
  \end{tikzpicture}
  \caption{Robustness of oscillatory and excitable networks ($\netN=4$) to parameter heterogeneity. The network is initialized with varying percentages $h$ of relative Gaussian parameter variations (varying $\gNa$, $\gK$, and $\tauDecay$ over $400$ trials). A valid outcome requires a strict, sustained, sequential spiking pattern ($n_0 \to n_1 \to n_2 \to n_3 \to n_4$). Solid lines represent the mean percentage of valid trials, while shaded regions indicate the 95\% confidence interval. The oscillatory network quickly breaks down when subjected to parameter variations, whereas the excitable network tolerates much larger changes.}
  \label{fig:robustness_sweep}
\end{figure}
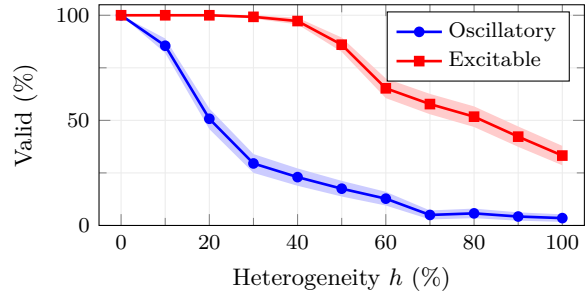

        To quantify this robustness, we apply relative Gaussian parameter variations $p_i \sim \mathcal{N}(\bar{p}_i, (h \cdot \bar{p}_i)^2)$ to key network parameters ($\gNa$, $\gK$, and $\tauDecay$). 
        We simulate $400$ trials for each heterogeneity level $h$, classifying a run as successful only if it yields a sustained, sequential pattern in the predetermined firing order. 
        Figure~\ref{fig:robustness_sweep} shows the fraction of these successful trials for both architectures.
        It demonstrates that, although the underlying neuronal and synaptic dynamics are (nearly) identical for the two architectures, the oscillatory network quickly exhibits pronounced fragility for heterogeneity levels above $h>10\%$.
        On the other hand, the excitable network maintains its functional rhythm across a much wider range of $h$ before its validity eventually deteriorates. 
        
        This highlights a core advantage of the event-driven approach: timing differences remain localised and do not interact across the network, since pattern generation relies on threshold-based event propagation rather than continuous phase-matching.
        
    \subsection{Local modulation}\label{sec:2:local-modulation}

        We now investigate the impact of local parameter modifications on the resulting global activity in both architectures.
        While the spatio-temporal dynamics in each case are governed by the total network period $\netT$ and the internal phase profile $\netPhaseProf$, the mechanisms translating local changes into global modulation differ fundamentally.
        
        In an excitable network, activity propagates strictly through discrete nodal delays $\nodeDelay{i}$. 
        The overall network period is simply the sum of these delays, $\netT = \sum_{i=0}^{\netN-1} \nodeDelay{i}$, from which the internal phase profile $\netPhaseProf = \{\netPhase_i\}_{i=0}^{\netN-1}$ follows directly as the set of relative delays, defined for each node by $\netPhase_i = \nodeDelay{i}/\netT$.
        In an oscillatory network, however, each node possesses an intrinsic natural frequency, making the emergent global rhythm a complex product of coupled interactions.

\begin{figure}[htb]
  \centering

  \begin{tikzpicture}
    \pgfplotsset{
      neuron scatter/.style={
        scatter,
        scatter src=explicit symbolic,
        point meta=\thisrow{neuron},
        scatter/classes={
          0={node0color},
          1={node1color},
          2={node2color},
          3={node3color}
        }
      }
    }
  
    \begin{groupplot}[
      group style={
        group size=2 by 2,
        horizontal sep=1.2cm,
        vertical sep=.7cm, 
        x descriptions at=edge bottom,
        y descriptions at=edge left,
      },
      width=7cm,
      height=3cm,
      enlarge x limits=0.05,
      restrict x to domain=1500:2500,  
      unbounded coords=jump,
      xlabel={Time (ms)},
      xmin=1500, xmax=2500,
      tick label style={font=\footnotesize},
      label style={font=\small},
      title style={font=\small\bfseries},
      filter discard warning=false,
      xtick={1500, 2000, 2500},
      every axis/.append style={
        grid=none,
        grid style={gray!15},
        minor tick num=0,
        yticklabel style={text width=1.5em, align=right},
        ylabel shift=6pt,
        ylabel style={align=center},
      },
      cycle list={
        {node0color},
        {node1color},
        {node2color},
        {node3color},
      },
    ]


    \nextgroupplot[
      title={Ring of excitable nodes},
      ylabel={Neuron},
      ymin=-1,
      ymax=4,
      ytick={0,1,2,3},
    ]
    \addplot[
      only marks,
      mark=|,
      mark size=5pt,
      line width=0.8pt,
      neuron scatter,
    ]
    table[
      x=time,
      y=neuron,
      col sep=comma,
    ] {plotdata/plotdata_excitable_raster.csv};

    \nextgroupplot[
      title={Equivalent network of oscillators},
      ymin=-1,
      ymax=4,
      ytick={0,1,2,3},    
      grid=none,
    ]
    \addplot[
      only marks,
      mark=|,
      mark size=5pt,
      line width=0.8pt,
      neuron scatter,
    ]
    table[
      x=time,
      y=neuron,
      col sep=comma,
    ] {plotdata/plotdata_oscillatory_raster.csv};


    \nextgroupplot[
      ylabel={Delay (ms)},
      ymin=0, ymax=50,
      grid=none,
      line width=0.5pt,
    ]
    \addplot+[line width=0.5pt, mark=|, mark size=1.5pt, mark repeat=1, x filter/.expression={(\thisrow{neuron} == 0 && x >= 1500 && x <= 2500) ? x : nan}] table[x=time, y=delay_ms, col sep=comma] {plotdata/plotdata_excitable_delay.csv};
    \addplot+[mark=|, mark size=1.5pt, mark repeat=1, x filter/.expression={(\thisrow{neuron} == 1 && x >= 1500 && x <= 2500) ? x : nan}] table[x=time, y=delay_ms, col sep=comma] {plotdata/plotdata_excitable_delay.csv};
    \addplot+[mark=|, mark size=1.5pt, mark repeat=1, x filter/.expression={(\thisrow{neuron} == 2 && x >= 1500 && x <= 2500) ? x : nan}] table[x=time, y=delay_ms, col sep=comma] {plotdata/plotdata_excitable_delay.csv};
    \addplot+[mark=|, mark size=1.5pt, mark repeat=1, x filter/.expression={(\thisrow{neuron} == 3 && x >= 1500 && x <= 2500) ? x : nan}] table[x=time, y=delay_ms, col sep=comma] {plotdata/plotdata_excitable_delay.csv};

    \nextgroupplot[
      ymin=0, ymax=50,
      grid=none,
    ]
    \addplot+[mark=|, mark size=1.5pt, mark repeat=1, x filter/.expression={(\thisrow{neuron} == 0 && x >= 1500 && x <= 2500) ? x : nan}] table[x=time, y=delay_ms, col sep=comma] {plotdata/plotdata_oscillatory_delay.csv};
    \addplot+[mark=|, mark size=1.5pt, mark repeat=1, x filter/.expression={(\thisrow{neuron} == 1 && x >= 1500 && x <= 2500) ? x : nan}] table[x=time, y=delay_ms, col sep=comma] {plotdata/plotdata_oscillatory_delay.csv};
    \addplot+[mark=|, mark size=1.5pt, mark repeat=1, x filter/.expression={(\thisrow{neuron} == 2 && x >= 1500 && x <= 2500) ? x : nan}] table[x=time, y=delay_ms, col sep=comma] {plotdata/plotdata_oscillatory_delay.csv};
    \addplot+[mark=|, mark size=1.5pt, mark repeat=1, x filter/.expression={(\thisrow{neuron} == 3 && x >= 1500 && x <= 2500) ? x : nan}] table[x=time, y=delay_ms, col sep=comma] {plotdata/plotdata_oscillatory_delay.csv};

    \end{groupplot}

    \node[anchor=south west, font=\small\bfseries\sffamily, inner sep=0pt] 
      at ([yshift=4pt,xshift=-15pt]group c1r1.north west) {(a)};
    \node[anchor=south west, font=\small\bfseries\sffamily, inner sep=0pt] 
      at ([yshift=4pt,xshift=-15pt]group c2r1.north west) {(b)};
    \node[anchor=south west, font=\small\bfseries\sffamily, inner sep=0pt] 
      at ([yshift=4pt,xshift=-15pt]group c1r2.north west) {(c)};
    \node[anchor=south west, font=\small\bfseries\sffamily, inner sep=0pt] 
      at ([yshift=4pt,xshift=-15pt]group c2r2.north west) {(d)};

    \setlength{\tabcolsep}{3pt}
    \node[
      draw=black!80,
      fill=white,
      font=\footnotesize,
      anchor=north,
      inner sep=2pt
    ] at ($(group c1r2.south)!0.5!(group c2r2.south) - (0, 0.6cm)$) {
      \begin{tabular}{*{4}{c}}
        \tikz[baseline=-0.5ex]\draw[node0color, thick, mark=*, mark size=1.5pt] (0,0) -- (0.4,0); $n_0$ &
        \tikz[baseline=-0.5ex]\draw[node1color, thick, mark=square*, mark size=1.5pt] (0,0) -- (0.4,0); $n_1$ &
        \tikz[baseline=-0.5ex]\draw[node2color, thick, mark=triangle*, mark size=1.5pt] (0,0) -- (0.4,0); $n_2$ &
        \tikz[baseline=-0.5ex]\draw[node3color, thick, mark=diamond*, mark size=1.5pt] (0,0) -- (0.4,0); $n_3$
      \end{tabular}
    };
  \end{tikzpicture}
  \caption{Single-synapse modulation induces global changes in oscillatory networks but remains localized in excitable networks ($\netN=4$). \textbf{(a), (b)} Raster plots and \textbf{(c), (d)} effective inter-neuron delays (i.e., the delay between node $\node{i-1}$ and $\node{i}$) contrast the network responses to an increase in $\tauDecay$ between nodes $\node{0}$ and $\node{1}$ at $t=2000\:\mathrm{s}$, where $\tauDecay=3.2\:\mathrm{ms}$ is increased to $9.5\:\mathrm{ms}$ and $4\:\mathrm{ms}$ for the excitable and oscillatory regimes, respectively. \textbf{(a), (c)} In the excitable network, the identical modulation alters only local dynamics, leaving the broader network unaffected. \textbf{(b), (d)} Conversely, in the oscillatory network, the perturbation propagates to alter all effective delays globally in a complex manner.}
  \label{fig:motivating_example}
\end{figure}
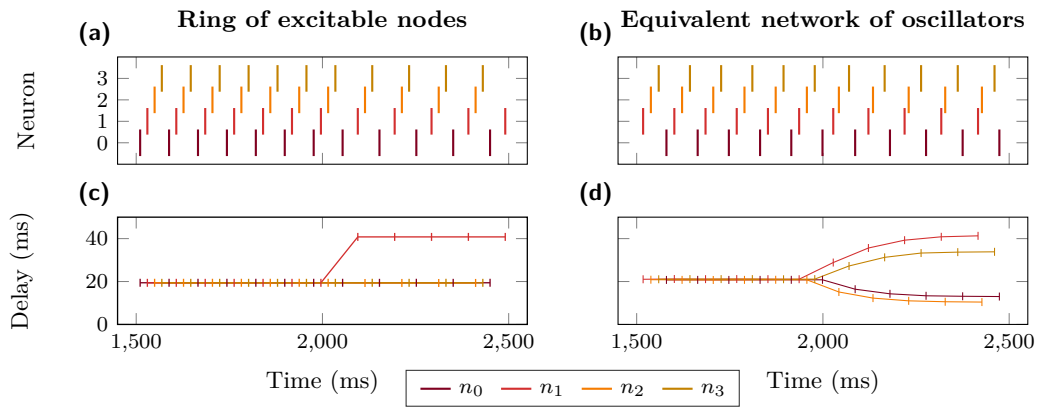

        Figure~\ref{fig:motivating_example} demonstrates this crucial difference through a simple experiment, contrasting spike raster plots (Figures~\ref{fig:motivating_example}(a), (b)) and effective inter-neuron delays (Figures~\ref{fig:motivating_example}(c), (d)): at $t = 2000\,\text{ms}$, we modify the network's local behaviour by increasing the synaptic time constant $\tauDecay$ from node $\node{0}$ to node $\node{1}$ in order to lengthen the delay between them from $20\,\text{ms}$ to $40\,\text{ms}$.
        Within the clock rhythm framework, this operation directly corresponds to enlarging one specific interval without intending to alter the remaining sequence.
        
        In the excitable network (Figures~\ref{fig:motivating_example}(a), (c)), this modulation predictably alters \textit{only} the local delay between node $\node{0}$ and node $\node{1}$.
        The remaining nodal dynamics are unaffected, proving that excitable networks provide decoupled local control over timing. 
        In contrast, in the oscillatory network (Figures~\ref{fig:motivating_example}(b), (d)), altering this single connection breaks the fragile phase-locking balance, causing unintended shifts in the effective delays across all nodes.
        Remember that, as discussed above, large levels of modulation may quickly result in the destruction of the rhythm in the oscillatory network.

        Note that in (certain) coupled oscillator networks, a desired phase profile tuning can be achieved by accounting for the global effects of the synaptic connections~\cite{seoCPGbasedControlTurtlelike2010}. 
        However, it is the locality of the tuning in the excitable network that we want to stress here. 
        Such local tuning is desirable in more complex networks or more complex modulation objectives.
                
    \subsection{Local analysis}
        These findings highlight why excitable networks offer a compelling alternative to networks of coupled oscillators: robust global coordination emerges from strictly local, nodal input--output characteristics. 
        Because rhythm generation relies on decoupled event propagation, the global network dynamics collapse into a sequence of isolated, local transitions.
        This locality eliminates the need to analyse complex, high-dimensional coupled systems; instead, global network behaviour can be rigorously analysed, predicted, and designed purely through the input--output response of individual nodes. 
        This principle of local analysis forms the primary theme for the remainder of this paper, establishing a fully distributed framework to analyse, design and tune event-driven rhythms.

\section{Endogenous rhythm generation}\label{sec:endogenous}

    In this section, we address the design problem for generating a rhythm \textit{in isolation}, that is, in a network that does not interact with its environment. 
    We postpone the interaction properties to the next section. 

 \subsection{Mechanism of rhythm generation}\label{sec:3:mechanism}
            We consider the design of rhythmic patterns in a network of excitable neurons interconnected by synapses. 
            Excitable neurons have the ability to \textit{fire}, that is, emit a voltage spike event, when triggered by a synaptic current (which itself is elicited by a presynaptic spike event). 
            This is the core mechanism of event propagation in the network.
            If the presynaptic neuron is excitatory, the synaptic current depolarises the postsynaptic neuron, eliciting a spike if the depolarization is above threshold.
            If the presynaptic neuron is inhibitory, the synaptic current hyperpolarises the postsynaptic neuron, possibly eliciting an event when the inhibition is released.
            The latter phenomenon is called \textit{rebound} excitability \cite{hindmarshModelReboundBursting1994}.
            In the present paper, the primary focus is on event propagation in a network of inhibitory rebound neurons, as rebound propagation provides a robust mechanism to propagate events with adjustable delays, see e.g. \cite{huoWinnerTakesAllMechanismEvent2025}.  
            
            Beyond the two classical currents of a mixed feedback neuron (fast activation of an inward current and slow activation of an outward current), two additional specific bio-inspired internal currents are used to control specific properties of the neuron.
            First, the rebound property is decoupled from excitatory spike generation and is controlled by a low-threshold, fast activation, slow de-inactivation, inward current ($\iNaReb$) which is inspired by the low-threshold calcium current ($\iCaT$), whose role has been studied extensively in the literature~\cite{dethierPositiveFeedbackCellular2015,hindmarshModelReboundBursting1994,wangModelTtypeCalcium1991, engbersDistinctRolesIT2011a}.
            Secondly, the latency of the spike generation mechanism is controlled by a fast activation, slow inactivation, outward current ($\iA$-potassium), inspired by the biological A-type potassium current~\cite{rushPotassiumAcurrentLow1995}.

            The full modelling details of the nodal and synaptic dynamics are provided in Appendix~\ref{sec:app:models}.


    \subsection{The event describing function}
        \subsubsection{The event map and event describing function}
\begin{figure}[htb]
\centering
\begin{tikzpicture}[
    >=Stealth,
    neuron/.style={circle, draw=black, fill=red!20, minimum size=0.85cm, font=\small\sffamily, thick},
    ring link/.style={->, thick, shorten >=2pt, 
    shorten <=2pt},
    cut gap/.style={pos=0.5, sloped, fill=white, inner sep=0pt, minimum width=0.26cm, minimum height=0.42cm},
    cut line/.style={draw=red!80!black, line width=1.3pt},
    zoom box/.style={dashed, draw=red!70!black, line width=0.8pt, rounded corners=3pt},
    zoom ray/.style={dashed, draw=gray!70!black, line width=0.8pt},
    block/.style={rectangle, draw=black, thick, minimum width=1.6cm, minimum height=0.95cm, align=center, font=\footnotesize},
    circ/.style={circle, draw=black, thick, minimum size=1.3cm, align=center, font=\footnotesize}
]

    \begin{scope}[shift={(0,0)}]
        \def\ringRadius{1.35cm}

        \node[neuron] (r0) at (90:\ringRadius) {0};
        \node[neuron, fill=red!35] (r1) at (0:\ringRadius) {1};
        \node[neuron] (r2) at (270:\ringRadius) {2};
        \node[neuron] (r3) at (180:\ringRadius) {3};

        \draw[ring link] (r2) to[bend left=20] (r3);
        \draw[ring link] (r3) to[bend left=20] (r0);

        \draw[ring link] (r0) to[bend left=20] node[cut gap] (cutIn) {} (r1);
        \draw[cut line] (cutIn.north) -- (cutIn.south);

        \draw[ring link] (r1) to[bend left=20] node[cut gap] (cutOut) {} (r2);
        \draw[cut line] (cutOut.north) -- (cutOut.south);


        \draw[zoom ray] ($(r1)+(0.45, 0.5)$) -- ++(0.6, 0.3);
        \draw[zoom ray] ($(r1)+(0.45,-0.5)$) -- ++(0.6,-0.3);

    \end{scope}

    \begin{scope}[shift={(4.6,0)}]
        \node[block] (syn) at (0,0) {Synapse\\(E/A)};
        \node[circ, right=1cm of syn] (neu) {Excitable\\Neuron};
        \node[block, right=1cm of neu] (det) {Event\\Detector\\(A/E)};

        \draw[->, thick] (syn) -- node[above, font=\footnotesize] {$\iSyn(t)$} (neu);
        \draw[->, thick] (neu) -- node[above, font=\footnotesize] {$V(t)$} (det);

        \draw[thick, ->] (-2, 0) -- (syn.west);
        \foreach \x in {-1.9, -1.65, -1.4, -1.15} {
            \draw[thick] (\x, 0) -- (\x, 0.35);
        }

        \draw[thick, ->] (det.east) -- ++(1.2, 0);
        \foreach \x in {0.15, 0.4, 0.65, .9} {
            \draw[thick] ($(det.east)+(\x,0)$) -- ($(det.east)+(\x,0.35)$);
        }

    \end{scope}

\end{tikzpicture}
\caption{Ring architecture of 4 neurons (left) with a cut around node $\node{1}$, expanded into its constituent subsystems: an incoming synapse acting as an event-to-analog converter (E/A) from events to synaptic conductance, and excitable neuron converting synaptic conductance into membrane potential, and an event detection mechanism acting as an analog-to-event converter (A/E).}

\label{fig:ring_cut_expanded}
\end{figure}
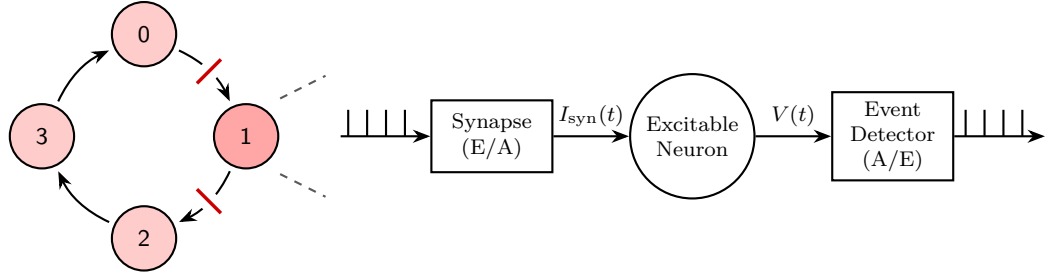

            In the elementary ring network of Figure~\ref{fig:ring}, each node acts as an adjustable delay between the pre- and post-synaptic events. 
            The event map that defines this delay is shown in  Figure~\ref{fig:ring_cut_expanded}: it consists of the series interconnection of the input synapse, modelled as an event-to-analog (E/A) converter; the neuron, modelled as a physical continuous-time circuit with input current $\iSyn(t)$ and output voltage $V(t)$; and the event readout, modelled as an analog-to-event (A/E) converter from the continuous-time signal $V(t)$ to a discrete event.
            
            Assuming a periodic input of presynaptic events, the event describing function ($\eDF$) characterises the nodal delay as a function of the input period $\inputPeriod$~\cite{wroblowskiDescribingFunctionsPhase2025}.
            This terminology is reminiscent from the describing function~\cite{khalilNonlinearSystems2002,wuFeedbackControlOscillations2014} of a continuous-time system, when analysing continuous-time limit cycle oscillations in feedback systems. 
            
            Consider a periodic input event sequence $\eventIn = \{\timeIn{k}\}_{k \in \mathbb{Z}}$ with period $\inputPeriod$ (such that $\timeIn{k+1} = \timeIn{k} + \inputPeriod$). 
            The describing event function assumes that the node operates in a $1{:}1$ phase-locked regime where each input event triggers exactly one output event (this is required for simple event propagation).
            In steady-state, the $\eDF$ maps $\eventIn$ to an output event sequence $\eventOut = \{\timeOut{k}\}_{k \in \mathbb{Z}}$ via a scalar delay $\nodeDelay{}$, such that:
            \begin{equation*}
                \timeOut{k} = \timeIn{k} + \delayFunc,
            \end{equation*}
            where $\delayFunc$ characterises the steady-state absolute nodal delay as a function of the input period.
            To express this timing relationship as a relative phase shift within the oscillation cycle, we may also define a \textit{relative} $\eDF$ as:
            \begin{equation*}
                \phaseFunc = \frac{\delayFunc}{\inputPeriod}.
            \end{equation*}

        \subsubsection{The $\eDF$ curve}
        
\begin{figure}[htb]
  \centering
  \begin{tikzpicture}
    \begin{groupplot}[
      group style={
        group size=2 by 1,
        horizontal sep=2.2cm,
      },
      scale only axis,
      width={0.33\textwidth}, 
      height=2.5cm,
      grid=none,
      tick label style={font=\small},
      label style={font=\small},
      ylabel style={align=center},
    ]

      \nextgroupplot[
        xlabel={$\inputPeriod$ (ms)},
        ylabel={$\delayFunc$ (ms)},
        yticklabel style={text width=4em, align=right},
        legend pos=north east,
        xmin=10, xmax=55,
        xtick={50},
        extra x ticks={\EdfTMin, \EdfTR},
        extra x tick labels={\shortstack{$\minPeriod$\\ \EdfTMin}, \shortstack{$\restPeriod$\\ \EdfTR}},
        extra x tick style={
          tick label style={font=\small, anchor=north}
        },
        ytick={\EdfDeltaInf, 12.4},
        yticklabels={
          \shortstack{$\delayInf$ \\ 11.66},
          \shortstack{$\Delta(\minPeriod)$ \\ 12.4}
        },
        ylabel shift=-6pt,
        legend style={
          font=\small,
          fill=white,
          fill opacity=0.8,
          draw opacity=1,
          text opacity=1
        }
      ]
        \addplot[
          color=blue!70!black,
          mark=none,
          line width=1.5pt,
          domain=10:50,
        ] table [
          x=period, 
          y=delay, 
          col sep=comma
        ] {plotdata/plotdata_edf_single_curve.csv};
        \addlegendentry{eDF $\delayFunc$}

        \addplot[
          color=red!80!black,
          dashed,
          line width=1pt,
          domain=10:55,
          samples=2
        ] {\EdfDeltaInf};
        \addlegendentry{$\delayInf$}

        \draw[gray, dashed, line width=1pt] 
          (axis cs:\EdfTMin, \pgfkeysvalueof{/pgfplots/ymin}) -- 
          (axis cs:\EdfTMin, \pgfkeysvalueof{/pgfplots/ymax});

        \draw[gray, dashed, line width=1pt] 
          (axis cs:\EdfTR, \pgfkeysvalueof{/pgfplots/ymin}) -- 
          (axis cs:\EdfTR, \pgfkeysvalueof{/pgfplots/ymax});

      \nextgroupplot[
        axis y line*=left,
        minor tick num=0,
        xlabel={$\inputPeriod$ (ms)},
        ylabel={Inhibitory $\delayFunc$ (ms)},
        ylabel shift=-24pt,
        y axis line style={blue!70!black},
        yticklabel style={font=\small, text=blue!70!black, text width=4em, align=right},
        ylabel style={font=\small, text=blue!70!black, align=center},
        xmin=5, xmax=55,
        ymin=11.5, ymax=12.6,
        xtick={10, 20, 30, 40, 50},
        ytick={11.6, 12.0, 12.4},
      ]
        \addplot[
          color=blue!70!black,
          mark=none,
          line width=1.5pt,
        ] table [
          x=period,
          y=delay,
          col sep=comma
        ] {plotdata/plotdata_edf_inh_curve.csv};
        \label{plot:edf_inh}
        
    \end{groupplot}

    \begin{axis}[
      at={(group c2r1.south west)}, 
      anchor=south west,
      scale only axis,
      width={0.33\textwidth},
      height=2.5cm,
      axis y line*=right,
      axis x line=none,
      ylabel={Excitatory $\delayFunc$ (ms)},
      ylabel shift=0pt,
      grid=none,
      tick label style={font=\small},
      label style={font=\small},
      y axis line style={orange!85!black},
      yticklabel style={font=\small, text=orange!85!black},
      ylabel style={font=\small, text=orange!85!black, align=center},
      xmin=5, xmax=55,
      ymin=0.6, ymax=1.9,
      ytick={0.8, 1.2, 1.6},
      legend pos=north east,
      legend style={
        font=\small,
        fill=white,
        fill opacity=0.8,
        draw opacity=1,
        text opacity=1
      }
    ]
      \addlegendimage{/pgfplots/refstyle=plot:edf_inh}
      \addlegendentry{Inhibitory eDF}

      \addplot[
        color=orange!85!black,
        mark=none,
        line width=1.5pt,
      ] table [
        x=period,
        y=delay,
        col sep=comma
      ] {plotdata/plotdata_edf_exc_curve.csv};
      \addlegendentry{Excitatory eDF}
    \end{axis}

    \node[anchor=south west, font=\small\bfseries\sffamily, inner sep=0pt] 
      at ([yshift=4pt, xshift=-15pt]group c1r1.north west) {(a)};
    \node[anchor=south west, font=\small\bfseries\sffamily, inner sep=0pt] 
      at ([yshift=4pt, xshift=-15pt]group c2r1.north west) {(b)};

  \end{tikzpicture}
  
  \caption{Event describing function (eDF) curve $\delayFunc$ for a single excitable neuron. \textbf{(a)} Single-neuron eDF curve and characteristic metrics. \textbf{(b)} Comparison of inhibitory (rebound) and excitatory (direct) eDF curves.}
  \label{fig:single_neuron_edf}
\end{figure}
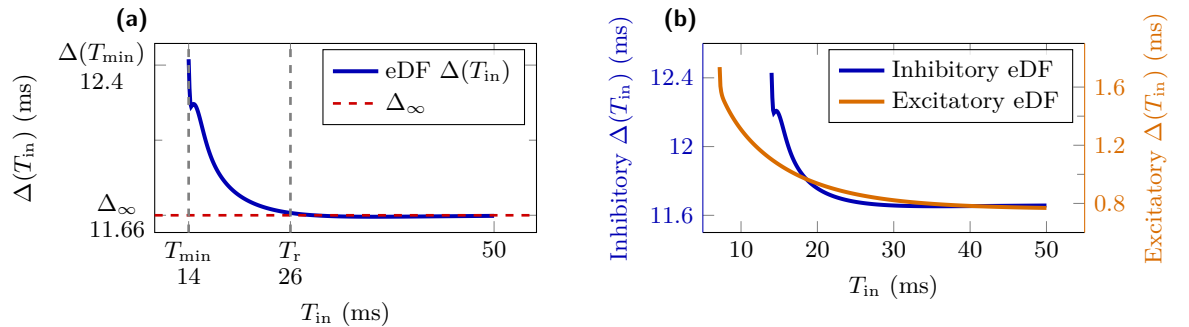

            The $\eDF$ curve $\delayFunc$ is constructed by numerically integrating the nodal continuous-time system dynamics across a sweep of input periods $\inputPeriod$, each time extracting the steady-state time shift $\nodeDelay{}$. 
            The resulting $\eDF$ curve reveals three distinct regimes, as shown in Figure~\ref{fig:single_neuron_edf}(a):    
            \input{plotdata/plotdata_edf_single_curve.tex}

            For large input periods ($\inputPeriod \ge \restPeriod$), the nodal delay is the same as in response to a single isolated input event. 
            This is due to the fading memory nature of the excitable neuron~\cite{cecconiContractionExcitableSystems2026}: its memory of the past decays over time, meaning that only the last input event determines the output when the input events are sparse enough. 
            For increasing input periods, the nodal delay converges to a constant baseline ($\delayFunc = \delayInf$) and the relative phase shift decays asymptotically toward zero ($\lim_{\inputPeriod \to \infty} \phaseFunc = 0$). 
            For small input periods, the  $1{:}1$ entrainment fails. 
            This can have various underlying causes: driving the system faster than $\minPeriod$ causes excitatory nodes to exhibit phase slipping or higher-order $N:M$ locking (firing $N$ times for every $M$ input cycles), while inhibitory nodes enter a state of sustained inhibition that completely suppresses spike generation.
            In the intermediate regime where $\minPeriod<T<\restPeriod$, the steady-state delay decreases to the asymptotic value $\Delta_\infty$ and may exhibit nonlinear effects.

            The existence of $\restPeriod$ and the asymptotic constant regime is an important consequence of nodal excitability. In practice, this allows the $\eDF$ to be computed from a single input event provided the global network period is large enough (i.e., $\netT>\restPeriod$). By ensuring the network operates in this fully-reset regime, the local nodal delay decouples entirely from the global period, remaining fixed at $\delayInf$. This yields a major computational advantage: instead of integrating the system dynamics over multiple cycles to reach steady state, the exact steady-state delay can be extracted from one isolated transient response.

            At the network level, determining whether an autonomous oscillation can emerge simply becomes an algebraic question. For instance, for a ring network, a global period $\netT>0$ must be a solution of the network period equation:
            \begin{equation*}
                \sum_i^{\netN} \netPhase_i(\netT) = 1.
            \end{equation*}

            From a design perspective, inhibition is the preferred mechanism of transmission because it provides a longer, highly controllable nodal delay with two distinct dimensions of tuning. 
            As shown in Figure~\ref{fig:single_neuron_edf}(b), the $\eDF$ behaves fundamentally differently depending on the synapse type. 
            Excitatory delays are inherently brief, restricted by the fast time constant of the spike upstroke and only the latency current ($\iA$) as control parameter. 
            In contrast, inhibitory delays are governed by both the synaptic decay time ($\tauDecay$) and the neuron's internal rebound latency ($\iA$). 
            This dual-control mechanism yields a broad tuning range, enabling even small networks to easily satisfy the global period condition.
    
    \subsection{Tuning the event map}

        The local analysis of the $\eDF$ establishes a principled foundation for network design. 
        To enable a design framework through local nodal tuning, it is enough to identify key parameters that control the nodal delay. 
        We make a distinction between internal and external tuning: internal tuning adjusts the rebound latency of the excitable neuron ($\iA$) whereas external tuning adjusts the synaptic time constant ($\tauDecay$).
        
        As introduced earlier, we exploit the well-known functionality of the $\mathrm{A}$-type potassium current ($\iA$) in order to tune the nodal delay internally. 
        The slow inactivation of this current provides a direct control of the rebound latency (that is, the transient behaviour after release of inhibition).  
        Tuning its maximal conductance ($\gA$) affords direct, exponential control over the event delay without compromising overall excitability.
        
\begin{figure}[htb]
  \centering
  \begin{tikzpicture}
    \begin{groupplot}[
      group style={
        group size=2 by 1,
        horizontal sep=2.0cm,
      },
      width={0.5\textwidth - 0.5cm},
      height=4.5cm,
      grid=none,
      unbounded coords=discard,
      tick label style={font=\small},
      xlabel style={font=\small, yshift=0.3em},
      ylabel style={font=\small},
      label style={font=\small},
      legend pos=north west,
      legend style={
        font=\tiny,
        fill=white,
        fill opacity=0.8,
        draw opacity=1,
        text opacity=1
      }
    ]

      \nextgroupplot[
        xlabel={$\tauDecay$ (ms)},
        ylabel={Period (ms)},
        xtick={1,5,10},
        xmin=1, xmax=10,
      ]
        \addplot[color=green!50!black, mark=*, mark size=1.8pt, line width=1pt] 
          table [x=param_val, y=t_min, col sep=comma] {plotdata/plotdata_metrics_sweep_edf_synapse_tau_decay.csv};
        \addlegendentry{$\minPeriod$}

        \addplot[color=red!70!black, mark=triangle*, mark size=2pt, line width=1pt] 
          table [x=param_val, y=delta_inf, col sep=comma] {plotdata/plotdata_metrics_sweep_edf_synapse_tau_decay.csv};
        \addlegendentry{$\delayInf$}

      \nextgroupplot[
        xmode=log,
        log basis x=10,
        x dir=reverse,  
        xlabel={$\gA$ (mS/cm$^2$)},
        ylabel={Period (ms)},
        xtick={1.6, 1.0, 0.3, 0.1, 0.01},
        xticklabels={$0.01$, $0.6$, $1.3$, $1.5$, $1.6$},
        log ticks with fixed point,
      ]
        \addplot[color=green!50!black, mark=*, mark size=1.8pt, line width=1pt] 
          table [x expr={1.65 - \thisrow{param_val}}, y=t_min, col sep=comma] {plotdata/plotdata_metrics_sweep_edf_neuron_g_a.csv};
        \addlegendentry{$\minPeriod$}

        \addplot[color=red!70!black, mark=triangle*, mark size=2pt, line width=1pt] 
          table [x expr={1.65 - \thisrow{param_val}}, y=delta_inf, col sep=comma] {plotdata/plotdata_metrics_sweep_edf_neuron_g_a.csv};
        \addlegendentry{$\delayInf$}

    \end{groupplot}

    \node[anchor=south west, font=\small\bfseries\sffamily, inner sep=0pt] 
      at ([yshift=4pt, xshift=-15pt]group c1r1.north west) {(a)};
    \node[anchor=south west, font=\small\bfseries\sffamily, inner sep=0pt] 
      at ([yshift=4pt, xshift=-15pt]group c2r1.north west) {(b)};

  \end{tikzpicture}
  
  \caption{Critical event describing function ($\eDF$) metrics ($\minPeriod$ and $\delayInf$) as functions of \textbf{(a)} synaptic decay time $\tauDecay$ and \textbf{(b)} A-type potassium conductance $\gA$.}
  \label{fig:edf_metrics_sweeps}
\end{figure}
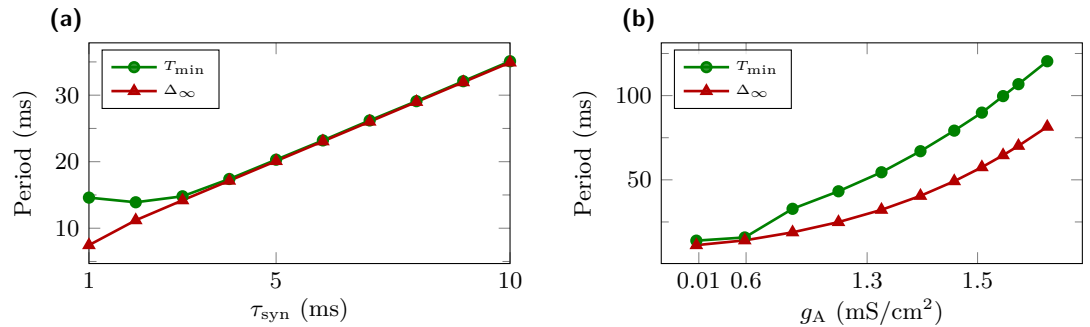

        Figure~\ref{fig:edf_metrics_sweeps} illustrates how $\tauDecay$ and $\gA$ provide complementary control parameters to shape the $\eDF$: the external, synaptic time constant $\tauDecay$ offers fine, linear tuning (panel~(a)), while the internal, cellular A-type conductance $\gA$ offers high-dynamic-range exponential tuning (panel~(b)).
        This design procedure offers a simple illustration of how the neurophysiology of excitability can be leveraged and serve as a biological inspiration for neuromorphic control.
                
    \subsection{Local tuning of network properties}
     
        \subsubsection{Tuning the global period}
            As a first illustration, we show how nodal parameters (the synaptic decay time constant $\tauDecay$ and intrinsic $\iA$-conductance $\gA$) can be exploited to tune the global network period $\netT$.
            In prior work, we showed that the collective network period can be predicted by summing individual nodal delays, and how $\gSyn$ and $\tauDecay$ shape the empirical $\eDF$ curve~\cite{wroblowskiDescribingFunctionsPhase2025}.
            Extending this principle to global network control, we evaluate two regimes: homogeneous modulation across all $\netN$ nodes, and localised heterogeneous modulation, where a single target node is adjusted while the remaining nodes remain fixed.
            
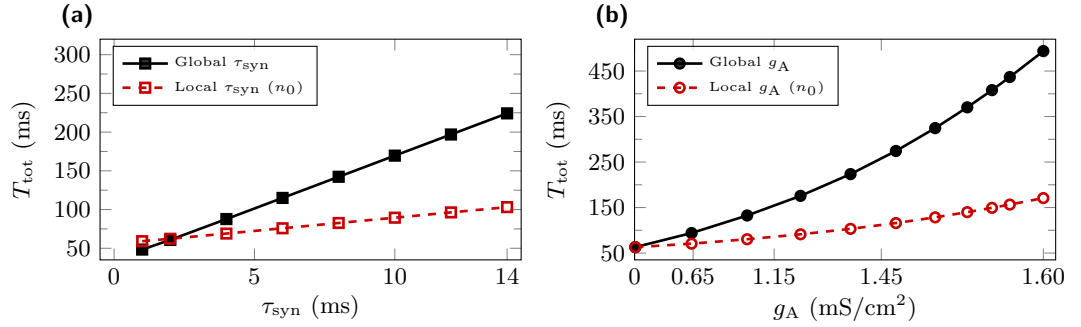
\begin{figure}[htb]
  \centering
  \begin{tikzpicture}
    \begin{groupplot}[
      group style={
        group size=2 by 1,
        horizontal sep=1.5cm,
      },
      width={0.5\textwidth - 0.5cm},
      height=4.5cm,
      grid=none,
      unbounded coords=discard,
      tick label style={font=\small},
      xlabel style={font=\small, yshift=0.3em},
      ylabel style={font=\small},
      label style={font=\small},
      title style={font=\small\bfseries},
      legend pos=north west,
      legend style={
        font=\tiny,
        fill=white,
        fill opacity=0.8,
        draw opacity=1,
        text opacity=1
      }
    ]

      \nextgroupplot[
        xlabel={$\tauDecay$ (ms)},
        ylabel={$\netT$ (ms)},
        xmin=-0.5, xmax=14.5,
        xtick={0,5,10,14},
        ymin=35, ymax=320,
        ytick={50,100,150,200,250,300},
      ]
        \addplot[color=black, mark=square*, mark size=1.8pt, line width=1pt] 
          table [x=param_val, y=network_period, col sep=comma] {plotdata/ring_sweep_synapse_tau_decay_all_nodes.csv};
        \addlegendentry{Global $\tauDecay$}

        \addplot[color=red!80!black, mark=square, dashed, mark size=1.8pt, line width=1pt, mark options={solid}] 
          table [x=param_val, y=network_period, col sep=comma] {plotdata/ring_sweep_synapse_tau_decay_node0.csv};
        \addlegendentry{Local $\tauDecay$ ($n_0$)}

      \nextgroupplot[
        xmode=log,
        log basis x=10,
        x dir=reverse,  
        xlabel={$\gA$ (mS/cm$^2$)},
        ylabel={$\netT$ (ms)},
        xmin=0.045, xmax=1.65,
        xtick={1.65, 1.0, 0.5, 0.2, 0.05},
        xticklabels={$0$, $0.65$, $1.15$, $1.45$, $1.60$},
        log ticks with fixed point,
        ymin=35, ymax=520,
        ytick={50,150,250,350,450},
      ]
        \addplot[color=black, mark=*, mark size=1.8pt, line width=1pt] 
          table [x expr={1.65 - \thisrow{param_val}}, y=network_period, col sep=comma] {plotdata/ring_sweep_neuron_g_a_all_nodes.csv};
        \addlegendentry{Global $\gA$}

        \addplot[color=red!80!black, mark=o, dashed, mark size=1.8pt, line width=1pt, mark options={solid}] 
          table [x expr={1.65 - \thisrow{param_val}}, y=network_period, col sep=comma] {plotdata/ring_sweep_neuron_g_a_node0.csv};
        \addlegendentry{Local $\gA$ ($n_0$)}

    \end{groupplot}

    \node[anchor=south west, font=\small\bfseries\sffamily, inner sep=0pt] 
      at ([yshift=4pt, xshift=-15pt]group c1r1.north west) {(a)};
    \node[anchor=south west, font=\small\bfseries\sffamily, inner sep=0pt] 
      at ([yshift=4pt, xshift=-15pt]group c2r1.north west) {(b)};

  \end{tikzpicture}
  \caption{Network period response ($\netN=4$) under homogeneous (all nodes) and heterogeneous modulation (node $\node{0}$). \textbf{(a)} Synaptic decay sweep ($\tauDecay$). \textbf{(b)} $\iA$-conductance sweep ($\gA$) plotted on a log-transformed axis ($z = 1.65 - \gA$, axis inverted). Nominal parameter values for heterogeneous modulation are $\tauDecay=2\:\mathrm{ms}$ and $\gA=0$.}
  \label{fig:tuning:period}
\end{figure}

            As shown in Figure~\ref{fig:tuning:period}, both homogeneous and heterogeneous adjustments of the synaptic time constant $\tauDecay$ and the intrinsic A-type conductance $\gA$ serve as effective control knobs for frequency regulation.
            Specifically, Figure~\ref{fig:tuning:period}(a) demonstrates that scaling $\tauDecay$ (both globally across all nodes and locally for a single target node) tunes the global network period $\netT$ in a smooth, nearly linear fashion.
            In contrast, Figure~\ref{fig:tuning:period}(b) shows that varying the A-type potassium conductance $\gA$ governs $\netT$ across a much wider dynamic range, exhibiting an exponential scaling behaviour.
            
\begin{figure}[htb]
  \centering
  \begin{tikzpicture}

    \begin{groupplot}[
      group style={
        group size=2 by 2,
        horizontal sep=1.2cm,
        vertical sep=1.2cm, 
        y descriptions at=edge left,
      },
      width=7cm,
      height=3cm,
      unbounded coords=jump,
      tick label style={font=\footnotesize},
      label style={font=\small},
      title style={font=\small\bfseries},
      filter discard warning=false,
      every axis/.append style={
        grid=none,
        minor tick num=0,
        tick align=inside,
        tick pos=left,
        yticklabel style={text width=1.8em, align=right}, 
        ylabel style={xshift=-4pt},                       
      },
    ]


      \nextgroupplot[
        title={Homogeneous},
        ylabel={Delay (ms)},
        xmin=0, xmax=1400,
        xtick={0,400,800,1200},
        ymin=10, ymax=30,
        ytick={10,15,20,25,30},
        line width=0.5pt,
      ]
      \addplot+[color=node0color, mark=|, mark size=1.5pt, mark repeat=1, x filter/.expression={\thisrow{neuron} == 0 ? x : nan}] table[x=time, y=delay_ms, col sep=comma] {plotdata/ring_network_delay.csv};
      \addplot+[color=node1color, mark=|, mark size=1.5pt, mark repeat=1, x filter/.expression={\thisrow{neuron} == 1 ? x : nan}] table[x=time, y=delay_ms, col sep=comma] {plotdata/ring_network_delay.csv};
      \addplot+[color=node2color, mark=|, mark size=1.5pt, mark repeat=1, x filter/.expression={\thisrow{neuron} == 2 ? x : nan}] table[x=time, y=delay_ms, col sep=comma] {plotdata/ring_network_delay.csv};
      \addplot+[color=node3color, mark=|, mark size=1.5pt, mark repeat=1, x filter/.expression={\thisrow{neuron} == 3 ? x : nan}] table[x=time, y=delay_ms, col sep=comma] {plotdata/ring_network_delay.csv};

      \nextgroupplot[
        title={Heterogeneous},
        xmin=0, xmax=1400,
        xtick={0,400,800,1200},
        ymin=10, ymax=30,
        ytick={10,15,20,25,30},
        line width=0.5pt,
      ]
      \addplot+[color=node0color, mark=|, mark size=1.5pt, mark repeat=1, x filter/.expression={\thisrow{neuron} == 0 ? x : nan}] table[x=time, y=delay_ms, col sep=comma] {plotdata/ring_network_delay_n0.csv};
      \addplot+[color=node1color, mark=|, mark size=1.5pt, mark repeat=1, x filter/.expression={\thisrow{neuron} == 1 ? x : nan}] table[x=time, y=delay_ms, col sep=comma] {plotdata/ring_network_delay_n0.csv};
      \addplot+[color=node2color, mark=|, mark size=1.5pt, mark repeat=1, x filter/.expression={\thisrow{neuron} == 2 ? x : nan}] table[x=time, y=delay_ms, col sep=comma] {plotdata/ring_network_delay_n0.csv};
      \addplot+[color=node3color, mark=|, mark size=1.5pt, mark repeat=1, x filter/.expression={\thisrow{neuron} == 3 ? x : nan}] table[x=time, y=delay_ms, col sep=comma] {plotdata/ring_network_delay_n0.csv};


      \nextgroupplot[
        ylabel={$V$ (mV)},
        xlabel={Time (ms)},
        xmin=500, xmax=1100,
        xtick={500,800,1100},
        restrict x to domain=500:1100,
        ymin=-25, ymax=105,
        ytick={0,50,100}, 
      ]
      \addplot[color=node0color, line width=0.8pt, each nth point=5] table[x=time, y=n0, col sep=comma] {plotdata/ring_network_voltage.csv};
      \addplot[color=node1color, line width=0.8pt, each nth point=5] table[x=time, y=n1, col sep=comma] {plotdata/ring_network_voltage.csv};
      \addplot[color=node2color, line width=0.8pt, each nth point=5] table[x=time, y=n2, col sep=comma] {plotdata/ring_network_voltage.csv};
      \addplot[color=node3color, line width=0.8pt, each nth point=5] table[x=time, y=n3, col sep=comma] {plotdata/ring_network_voltage.csv};

      \nextgroupplot[
        xlabel={Time (ms)},
        xmin=500, xmax=1100,
        xtick={500,800,1100},
        restrict x to domain=500:1100,
        ymin=-25, ymax=105,
        ytick={0,50,100}, 
      ]
      \addplot[color=node0color, line width=0.8pt, each nth point=5] table[x=time, y=n0, col sep=comma] {plotdata/ring_network_voltage_n0.csv};
      \addplot[color=node1color, line width=0.8pt, each nth point=5] table[x=time, y=n1, col sep=comma] {plotdata/ring_network_voltage_n0.csv};
      \addplot[color=node2color, line width=0.8pt, each nth point=5] table[x=time, y=n2, col sep=comma] {plotdata/ring_network_voltage_n0.csv};
      \addplot[color=node3color, line width=0.8pt, each nth point=5] table[x=time, y=n3, col sep=comma] {plotdata/ring_network_voltage_n0.csv};

    \end{groupplot}

    \node[anchor=south west, font=\small\bfseries\sffamily, inner sep=0pt] 
      at ([yshift=6pt, xshift=-15pt]group c1r1.north west) {(a)};
    \node[anchor=south west, font=\small\bfseries\sffamily, inner sep=0pt] 
      at ([yshift=6pt, xshift=-15pt]group c2r1.north west) {(b)};
    \node[anchor=south west, font=\small\bfseries\sffamily, inner sep=0pt] 
      at ([yshift=6pt, xshift=-15pt]group c1r2.north west) {(c)};
    \node[anchor=south west, font=\small\bfseries\sffamily, inner sep=0pt] 
      at ([yshift=6pt, xshift=-15pt]group c2r2.north west) {(d)};

    \setlength{\tabcolsep}{3pt}
    \node[
      draw=black!80,
      fill=white,
      font=\footnotesize,
      anchor=north,
      inner sep=2pt
    ] at ($(group c1r2.south)!0.5!(group c2r2.south) - (0, 0.6cm)$) {
      \begin{tabular}{*{4}{c}}
        \tikz[baseline=-0.5ex]\draw[node0color, thick, mark=*, mark size=1.5pt] (0,0) -- (0.4,0); $n_0$ &
        \tikz[baseline=-0.5ex]\draw[node1color, thick, mark=square*, mark size=1.5pt] (0,0) -- (0.4,0); $n_1$ &
        \tikz[baseline=-0.5ex]\draw[node2color, thick, mark=triangle*, mark size=1.5pt] (0,0) -- (0.4,0); $n_2$ &
        \tikz[baseline=-0.5ex]\draw[node3color, thick, mark=diamond*, mark size=1.5pt] (0,0) -- (0.4,0); $n_3$
      \end{tabular}
    };

  \end{tikzpicture}
  \caption{Comparison of network dynamics under homogeneous (\textbf{a}, \textbf{c}) versus heterogeneous (\textbf{b}, \textbf{d}) modulation, triggered by increasing $\tauDecay$ from $2.15$ to $6$ at $t=700$. The heterogeneous scheme exclusively targets $\node{0}$. \textbf{(a)}, \textbf{(b)} Inter-neuron delay values. \textbf{(c)}, \textbf{(d)} Membrane potential traces.}
  \label{fig:tuning:period:example}
\end{figure}
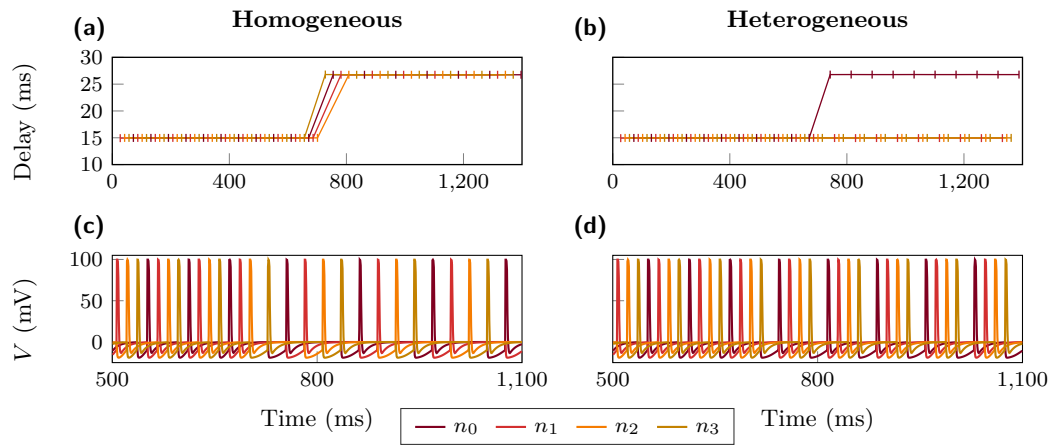

            Furthermore, the transitions between nominal and modulated rhythms occur cleanly without unexpected transient dynamics, as shown in Figure~\ref{fig:tuning:period:example}.
            Specifically, Figures~\ref{fig:tuning:period:example}(a) and (b) illustrate the smooth temporal evolution of inter-neuron delays under homogeneous and heterogeneous parameter updates, respectively (changing $\tauDecay$ from $2.15$ to $6$ at $t=700$).
            At the same time, the corresponding membrane potential traces in Figures~\ref{fig:tuning:period:example}(c) and (d) confirm that spike generation remains phase-locked and stable throughout the transition, avoiding skipped cycles or bursting anomalies.
            This rapid, stable adaptation stems directly from the event-propagation architecture, establishing $\tauDecay$ and $\gA$ as robust control knobs for network period regulation.

    
        \subsubsection{Tuning the phase profile}
            A complete description of the network rhythm requires both its global period and its phase profile $\netPhaseProf$ (the set of relative nodal delays $\nodeDelayRel{i}$). 
            Because these delays operate independently, modulating individual event delays $\nodeDelay{i}$ allows us to seamlessly reshape the duty cycle without destabilizing rhythmic event generation.
            
            Because homogeneous nodes cannot generate asymmetric phase profiles, we consider heterogeneous tuning to achieve a desired phase profile $\netPhaseProf^*$ (e.g., representing a desired locomotor gait). 
            As an example, we study the transition from a uniform splayed walking gait with phase offsets $\netPhaseProf_{\mathrm{walk}}^{\mathrm{offsets}} = \{0, 0.25, 0.5, 0.75\}$ to a galloping gait with offsets $\netPhaseProf_{\mathrm{gallop}}^{\mathrm{offsets}} = \{0, 0.4, 0.5, 0.9\}$ in the case of a quadrupedal CPG ($N=4$ nodes corresponding to the four limbs \textit{LF}, \textit{LH}, \textit{RF}, and \textit{RH}). 
            Assuming an underlying phase--task mapping handles node reassignment during the gait transition to establish the correct sequential ordering, the remaining transition is achieved by reshaping the phase profile from $\netPhaseProf_{\mathrm{walk}} = \{0.25, 0.25, 0.25, 0.25\}$ to $\netPhaseProf_{\mathrm{gallop}} = \{0.1, 0.4, 0.1, 0.4\}$.
    
            Starting from the nominal walking gait ($a$) where $\nodeDelay{i}^a = \SI{40}{\milli\second}$, we consider two approaches to obtain the target galloping gait ($b$).
    
\begin{figure}[htb]
  \centering
  \begin{tikzpicture}

    \pgfplotsset{
      neuron scatter/.style={
        scatter,
        scatter src=explicit symbolic,
        point meta=\thisrow{neuron},
        scatter/classes={
          0={node0color},
          1={node1color},
          2={node2color},
          3={node3color}
        }
      }
    }

    \begin{groupplot}[
      group style={
        group size=2 by 3,
        horizontal sep=1.5cm,
        vertical sep=0.7cm,
        x descriptions at=edge bottom,
        y descriptions at=edge left,
      },
      width=6.3cm,
      height=2.8cm,
      enlarge x limits=0.05,
      restrict x to domain=1500:2500,
      unbounded coords=discard,
      xlabel={Time (ms)},
      xmin=1500, xmax=2500,
      tick label style={font=\footnotesize},
      label style={font=\small},
      title style={font=\small\bfseries},
      filter discard warning=false,
      every axis/.append style={
        grid=none,
        minor tick num=0,
        tick align=inside,
        tick pos=left,
        yticklabel style={text width=1.5em, align=right},
        ylabel shift=3pt,
        ylabel style={align=center},
      },
    ]


      \nextgroupplot[
        title={Fixed Period ($\netT$)},
        ylabel={Neuron},
        ymin=-0.6, ymax=3.6,
        ytick={0,1,2,3},
      ]
      \addplot[
        only marks,
        mark=|,
        mark size=5pt,
        line width=0.8pt,
        neuron scatter,
      ]
      table[
        x=time,
        y=neuron,
        col sep=comma,
      ] {plotdata/plotdata_duty_modulation_approach1__raster.csv};

      \nextgroupplot[
        title={Fixed Node ($\nodeDelay{i}$)},
        ymin=-0.6, ymax=3.6,
        ytick={0,1,2,3},
      ]
      \addplot[
        only marks,
        mark=|,
        mark size=5pt,
        line width=0.8pt,
        neuron scatter,
      ]
      table[
        x=time,
        y=neuron,
        col sep=comma,
      ] {plotdata/plotdata_duty_modulation_approach2__raster.csv};


      \nextgroupplot[
        ylabel={Delay (ms)},
        ymin=0, ymax=80,
        ytick={0,20,40,60},
        line width=0.5pt,
      ]
      \addplot+[color=node0color, mark=|, mark size=1.5pt, mark repeat=1, x filter/.expression={\thisrow{neuron} == 0 ? x : nan}] table[x=time, y=delay_ms, col sep=comma] {plotdata/plotdata_duty_modulation_approach1__delay.csv};
      \addplot+[color=node1color, mark=|, mark size=1.5pt, mark repeat=1, x filter/.expression={\thisrow{neuron} == 1 ? x : nan}] table[x=time, y=delay_ms, col sep=comma] {plotdata/plotdata_duty_modulation_approach1__delay.csv};
      \addplot+[color=node2color, mark=|, mark size=1.5pt, mark repeat=1, x filter/.expression={\thisrow{neuron} == 2 ? x : nan}] table[x=time, y=delay_ms, col sep=comma] {plotdata/plotdata_duty_modulation_approach1__delay.csv};
      \addplot+[color=node3color, mark=|, mark size=1.5pt, mark repeat=1, x filter/.expression={\thisrow{neuron} == 3 ? x : nan}] table[x=time, y=delay_ms, col sep=comma] {plotdata/plotdata_duty_modulation_approach1__delay.csv};

      \nextgroupplot[
        ymin=0, ymax=80,
        ytick={0,20,40,60},
        line width=0.5pt,
      ]
      \addplot+[color=node0color, mark=|, mark size=1.5pt, mark repeat=1, x filter/.expression={\thisrow{neuron} == 0 ? x : nan}] table[x=time, y=delay_ms, col sep=comma] {plotdata/plotdata_duty_modulation_approach2__delay.csv};
      \addplot+[color=node1color, mark=|, mark size=1.5pt, mark repeat=1, x filter/.expression={\thisrow{neuron} == 1 ? x : nan}] table[x=time, y=delay_ms, col sep=comma] {plotdata/plotdata_duty_modulation_approach2__delay.csv};
      \addplot+[color=node2color, mark=|, mark size=1.5pt, mark repeat=1, x filter/.expression={\thisrow{neuron} == 2 ? x : nan}] table[x=time, y=delay_ms, col sep=comma] {plotdata/plotdata_duty_modulation_approach2__delay.csv};
      \addplot+[color=node3color, mark=|, mark size=1.5pt, mark repeat=1, x filter/.expression={\thisrow{neuron} == 3 ? x : nan}] table[x=time, y=delay_ms, col sep=comma] {plotdata/plotdata_duty_modulation_approach2__delay.csv};


      \nextgroupplot[
        ylabel={Rel. Delay},
        ymin=0, ymax=0.5,
        ytick={0,0.1,0.25,0.4},
        line width=0.5pt,
      ]
      \addplot+[color=node0color, mark=|, mark size=1.5pt, mark repeat=1, x filter/.expression={\thisrow{neuron} == 0 ? x : nan}] table[x=time, y=rel_delay, col sep=comma] {plotdata/plotdata_duty_modulation_approach1__delay.csv};
      \addplot+[color=node1color, mark=|, mark size=1.5pt, mark repeat=1, x filter/.expression={\thisrow{neuron} == 1 ? x : nan}] table[x=time, y=rel_delay, col sep=comma] {plotdata/plotdata_duty_modulation_approach1__delay.csv};
      \addplot+[color=node2color, mark=|, mark size=1.5pt, mark repeat=1, x filter/.expression={\thisrow{neuron} == 2 ? x : nan}] table[x=time, y=rel_delay, col sep=comma] {plotdata/plotdata_duty_modulation_approach1__delay.csv};
      \addplot+[color=node3color, mark=|, mark size=1.5pt, mark repeat=1, x filter/.expression={\thisrow{neuron} == 3 ? x : nan}] table[x=time, y=rel_delay, col sep=comma] {plotdata/plotdata_duty_modulation_approach1__delay.csv};

      \nextgroupplot[
        ymin=0, ymax=0.5,
        ytick={0,0.1,0.25,0.4},
        line width=0.5pt,
      ]
      \addplot+[color=node0color, mark=|, mark size=1.5pt, mark repeat=1, x filter/.expression={\thisrow{neuron} == 0 ? x : nan}] table[x=time, y=rel_delay, col sep=comma] {plotdata/plotdata_duty_modulation_approach2__delay.csv};
      \addplot+[color=node1color, mark=|, mark size=1.5pt, mark repeat=1, x filter/.expression={\thisrow{neuron} == 1 ? x : nan}] table[x=time, y=rel_delay, col sep=comma] {plotdata/plotdata_duty_modulation_approach2__delay.csv};
      \addplot+[color=node2color, mark=|, mark size=1.5pt, mark repeat=1, x filter/.expression={\thisrow{neuron} == 2 ? x : nan}] table[x=time, y=rel_delay, col sep=comma] {plotdata/plotdata_duty_modulation_approach2__delay.csv};
      \addplot+[color=node3color, mark=|, mark size=1.5pt, mark repeat=1, x filter/.expression={\thisrow{neuron} == 3 ? x : nan}] table[x=time, y=rel_delay, col sep=comma] {plotdata/plotdata_duty_modulation_approach2__delay.csv};

    \end{groupplot}

    \node[anchor=south west, font=\small\bfseries\sffamily, inner sep=0pt] 
      at ([yshift=2pt, xshift=-15pt]group c1r1.north west) {(a)};
    \node[anchor=south west, font=\small\bfseries\sffamily, inner sep=0pt] 
      at ([yshift=2pt, xshift=-15pt]group c2r1.north west) {(b)};
    \node[anchor=south west, font=\small\bfseries\sffamily, inner sep=0pt] 
      at ([yshift=2pt, xshift=-15pt]group c1r2.north west) {(c)};
    \node[anchor=south west, font=\small\bfseries\sffamily, inner sep=0pt] 
      at ([yshift=2pt, xshift=-15pt]group c2r2.north west) {(d)};
    \node[anchor=south west, font=\small\bfseries\sffamily, inner sep=0pt] 
      at ([yshift=2pt, xshift=-15pt]group c1r3.north west) {(e)};
    \node[anchor=south west, font=\small\bfseries\sffamily, inner sep=0pt] 
      at ([yshift=2pt, xshift=-15pt]group c2r3.north west) {(f)};

    \setlength{\tabcolsep}{3pt}
    \node[
      draw=black!80,
      fill=white,
      font=\footnotesize,
      anchor=north,
      inner sep=2pt
    ] at ($(group c1r3.south)!0.5!(group c2r3.south) - (0, 0.6cm)$) {
      \begin{tabular}{*{4}{c}}
        \tikz[baseline=-0.5ex]\draw[node0color, thick, mark=*, mark size=1.5pt] (0,0) -- (0.4,0); $n_0$ &
        \tikz[baseline=-0.5ex]\draw[node1color, thick, mark=square*, mark size=1.5pt] (0,0) -- (0.4,0); $n_1$ &
        \tikz[baseline=-0.5ex]\draw[node2color, thick, mark=triangle*, mark size=1.5pt] (0,0) -- (0.4,0); $n_2$ &
        \tikz[baseline=-0.5ex]\draw[node3color, thick, mark=diamond*, mark size=1.5pt] (0,0) -- (0.4,0); $n_3$
      \end{tabular}
    };

  \end{tikzpicture}
  \caption{Comparison of phase profile modulation strategies in a 4-node network, with parameter changes applied at $t = 2000$. \textbf{(a)}, \textbf{(b)} Spike raster plots. \textbf{(c)}, \textbf{(d)} Absolute nodal delays ($\nodeDelay{i}$). \textbf{(e)}, \textbf{(f)} Relative nodal delays (normalized to total cycle period). The left column illustrates global modulation under a fixed total period $\netT$, where $g_a$ is shifted from $1.2$ to $0.0$ for $\node{0}$ and $\node{2}$, and to $1.45$ for $\node{1}$ and $\node{3}$, while incoming $\tauDecay$ for synapses $3 \to 0$ and $1 \to 2$ increases from $2.15$ to $2.4$. The right column shows local modulation preserving specific node delays ($\node{1}$ and $\node{3}$), achieved by dropping $g_a$ from $1.2$ to $0.0$ ($\node{0}$ and $\node{2}$) and decreasing $\tauDecay$ ($3 \to 0$ and $1 \to 2$) from $2.5$ to $0.5$. Apparent transients in relative phase stem from incremental variations in cycle period.}
  \label{fig:tuning:duty}
\end{figure}
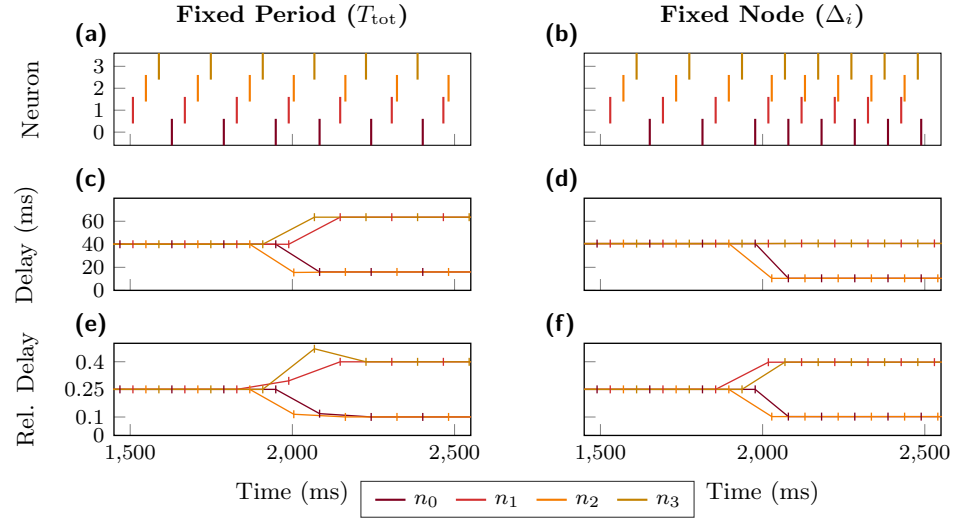

            In the first approach, we preserve the global network period ($\netT^a = \netT^b= \netT$) and derive the target absolute nodal delays directly from the phase profile: $\nodeDelay{i}^b = \netPhaseProf^*_i \netT$. 
            The result is shown in Figures~\ref{fig:tuning:duty}(a), (c), (e), where $\netT$ remains fixed while the gait rhythm is modulated at $t=2000$ (by dropping $\gA$ from $1.2$ to $0.0$ for nodes $\node{0}$ and $\node{2}$, increasing $\gA$ to $1.45$ for nodes $\node{1}$ and $\node{3}$, and increasing $\tauDecay$ from $2.15$ to $2.4$ for synapses $3 \to 0$ and $1 \to 2$) from $\netPhaseProf_{\mathrm{walk}}$ to $\netPhaseProf_{\mathrm{gallop}}$ (depicting the spike raster plot, absolute nodal delays $\nodeDelay{i}$, and relative nodal delays $\nodeDelayRel{i}$, respectively).
    
            In the second approach, we keep the absolute delays of nodes $\node{1}$ and $\node{3}$ unaltered while modulating nodes $\node{0}$ and $\node{2}$ to obtain the target phase profile $\netPhaseProf^*$. 
            Using the constraint $\nodeDelay{i}^b = \nodeDelay{i}^a$ for $i \in \{1,3\}$, the new total period becomes $\netT^b = \sum_{i=0}^{N-1} \nodeDelay{i}^b = \nodeDelay{0}^b + \nodeDelay{2}^b + \nodeDelay{1}^a + \nodeDelay{3}^a$.
            Substituting the phase relationship $\nodeDelay{i}^b = \netT^b \:\netPhaseProf^*_i$ for $i \in \{0,2\}$ gives $\netT^b = \netT^b (\netPhaseProf^*_0 + \netPhaseProf^*_2) + \nodeDelay{1}^a + \nodeDelay{3}^a$. 
            Finally, we solve for the updated nodal delays:
            \begin{equation*}
                \nodeDelay{i}^b = \frac{\netPhaseProf^*_i (\nodeDelay{1}^a + \nodeDelay{3}^a)}{1 - \netPhaseProf^*_0 - \netPhaseProf^*_2} \quad \text{for}\: i \in \{0,2\}.
            \end{equation*}
            
            The second strategy is illustrated in Figures~\ref{fig:tuning:duty}(b), (d), (f), where the delays of nodes $\node{0}$ and $\node{2}$ are shortened relative to the total period to achieve the desired profile $\netPhaseProf^*$ (implemented at $t=2000$ by dropping $\gA$ from $1.2$ to $0.0$ for nodes $\node{0}$ and $\node{2}$, and decreasing $\tauDecay$ from $2.5$ to $0.5$ for synapses $3 \to 0$ and $1 \to 2$).
            
            These two example highlight how a few local parameters enable a simple and intuitive tuning of the network rhythm properties.
        
        \subsubsection{Modular design and phase clustering}
        \label{sec:3:clustering}
            A basic ring network naturally enforces a strict, non-overlapping sequence of events. 
            To encode more sophisticated rhythmic behaviours, we can expand the spatial resolution of the rhythm simply by increasing the number of nodes $N$. 
            We illustrate the role of this higher-dimensional architecture using the quadrupedal locomotor gait from the previous section.

            As before, the simplest realisation is a $1{:}1$ phase--task mapping using $\netN=4$ nodes, where each node is assigned to one of the four limbs as before (\textit{LF}, \textit{LH}, \textit{RF}, and \textit{RH}). 
            In this setup, a spike from a node triggers the onset of that limb's active phase (e.g., the stance phase), and the delay to the subsequent node dictates the timing of the next limb's activation.
            While simple, a strict $1{:}1$ mapping implicitly enforces sequential, non-overlapping active states: the onset of one limb's phase perfectly coincides with the termination of the previous one.
            
            However, many locomotor gaits require muscle activations to overlap, or require the duration of a limb's active phase to be tuned independently of the broader gait sequence. 
            Because an excitable node outputs a single discrete event, a $1{:}1$ mapping provides only one timing reference (e.g., task onset) per limb. 
            To encode both the initiation and termination of a continuous action purely within the network's timing dynamics, we can expand the network resolution by assigning a \textit{cluster} of nodes to a single high-level task.
        
            For example, one might make use of a $2{:}1$ mapping, where two dedicated nodes govern a single task. 
            The first node's spike initiates the active state (e.g., muscle contraction), and the second node's spike terminates it. 
            By exposing the task duration as a tunable nodal delay, this clustering strategy decouples the duration of the activity from the delay to the subsequent task, granting independent control over the duty cycle of individual limbs. 
            
            If a gait additionally requires overlapping activations across multiple limbs, this can be achieved by defining the phase--task mapping such that a specific network phase explicitly represents a joint state (e.g., `both \textit{LF} and \textit{RH} are currently active').
            
            This modular way of clustering can be extended to arbitrary resolutions. 
            For example, the continuous motion of the \textit{LF} limb could be decomposed into a sequence of distinct functional sub-phases: initiation (heel-strike), maximum load (mid-stance), and release (toe-off). 
            By mapping a cluster of three nodes to this sequence, each sub-phase is exposed to local tuning.
            Consequently, a designer can selectively compress or stretch a specific sub-task (e.g., extending the stance duration to accommodate a heavier load) by tuning a single local delay, all while the event-driven architecture naturally preserves the integrity of the broader network rhythm.
    
    \subsection{Robustness and failure modes of the event map}
        The simplicity and robustness of the design framework provided by simple tools as the event Describing Function ($\eDF$) rely on acknowledging specific assumptions. 
        Most importantly, the assumption that the nodal input events are \textit{sparse} enough ($\netT>\minPeriod$), such that the nodal delay is well-defined and easy to compute. 
        The inhibitory nature of the network and the rebound property of the nodes are central to that assumption. 
        The design of the nodal rebound dynamics (determined by the intracellular current $\iNaReb$) is in that sense a critical ingredient for the robustness of the global network.
        Because the critical property is so localised, its robustness analysis is tractable, irrespective of the network size. 
        The key to robustness and tunability is to design the intracellular currents in such a way that the nodal delay is sensitive to few of them (for tunability) and insensitive to most of them (for robustness). 
        This design philosophy is a key inspiration from cellular neurophysiology. 


\section{Exogenous rhythm regulation}\label{sec:exogenous}

    Practical rhythmic networks are rarely designed to operate in isolation. 
    While they can generate a rhythm endogenously, they must also integrate sensory feedback from the external environment. 
    Such exogenous interaction is critical for robust operation; for example, it allows a locomotor network to seamlessly adapt its phase to uneven terrain. 
    This section introduces a second local analysis tool for this \textit{external} analysis.

    \subsection{The event phase response}\label{subsec:event_phase_response}

        In the classical theory of networks of oscillators, the key analysis tool for external properties is the Phase Response Curve (PRC)~\cite{schultheissPhaseResponseCurves2012, nakaoPhaseReductionApproach2016}. 
        We introduce the event Phase Response Curve ($\ePRC$) as the analog tool for a node that is excitable rather than oscillatory: starting from a nominal periodic regime, it measures the steady-state time shift induced by an additional perturbing periodic event sequence.
        
    \subsubsection{Definition}
    
        Starting from a nominal periodic regime (e.g. the outcome of an endogenous rhythm design such as in the previous section), we introduce an external input via an external input synapse. 
        The underlying assumption is that a neural network interacts with its environment via synaptic currents, consistent with the dynamics of biological neural circuits. 
        For the sake of illustration, we restrict the analysis to a nominal internal rhythm designed from an inhibitory network, as in the previous section.
        
        Let $\nominalT$ denote the period of the nominal, unperturbed endogenous input event sequence driving the excitable node into a $1{:}1$ phase-locked event oscillation. 
        A periodic perturbing event sequence $\eventP= \{\timeP{k}\}_{k \in \mathbb{Z}}$ is introduced at a time offset $\phaseOffset \in [0, \nominalT)$ within each nominal period through the external synapse, such that $\timeP{k} = \timeIn{k}+\phaseOffset$.
        We apply periodic rather than isolated perturbations to enable a stationary analysis, as in the previous section.

        The output event time shift $\phaseShift$ defines the event phase response:
        \begin{equation*}
            \ePRC(\phaseOffset) = \phaseShift = \timeOut{\mathrm{p}}(\phaseOffset) - \timeOut{\mathrm{nom}},
        \end{equation*}
        where $\timeOut{\mathrm{nom}}$ and $\timeOut{p}$ represent the nominal and perturbation-shifted output event timings, respectively. 
        
        When $\netT$ is sufficiently large such that the system fully relaxes to its resting state between each input event ($\netT > \restPeriod$), the $\ePRC$ can be safely and efficiently evaluated from a single event transmission.
        Furthermore, in this regime, perturbations arriving after the output event produce no phase shift.
        We therefore restrict the evaluation window to perturbations preceding the nominal input event, such that $\phaseOffset \in [-\nominalT, \nodeDelay{\mathrm{nom}})$.

        Evaluating $\phaseShift$ across a sweep of time offsets $\phaseOffset$ constructs the complete event phase response curve for a given endogenous behaviour.
        In parallel to classical PRC analysis, this curve defines the system's capacity for entrainment under frequency detuning ($\perturbT \neq \nominalT$). 
        This relationship relies on three principles: first, to remain synchronized, the node must shift its timing by the exact period difference $\Delta T = \perturbT - \nominalT$ every cycle. 
        Second, vertically shifting the $\ePRC$ by this $\Delta T$ allows us to find the stable phase-locking offsets at the new zero-crossings. 
        Finally, the amplitude of the $\ePRC$ establishes the absolute entrainment bounds, as any detuning $\Delta T$ exceeding the curve's peak will prevent phase-locking entirely.

    \subsubsection{Contrast to phase response curves analysis}
    
        To place the $\ePRC$ within the established literature, we briefly contrast it with related phase reduction theories.
        
        Unlike the classical infinitesimal Phase Response Curve (iPRC), which evaluates direct state-trajectory pulses on autonomous limit-cycle oscillators, the $\ePRC$ characterises non-oscillatory, excitable systems driven by explicit synaptic dynamics.
        
        To account for such physiological synapses, the functional Phase Response Curve (fPRC) utilises finite-amplitude, repetitive perturbations~\cite{cuiFunctionalPhaseResponse2009}. 
        While the fPRC can successfully characterise non-oscillatory, rebound-excitable components, it operates in a \textit{closed-loop} fashion, adaptively applying perturbations at a fixed delay after each output spike. 
        In contrast, the $\ePRC$ evaluates responses to a strictly independent periodic sequence, yielding a purely open-loop, input--output mapping of forced phase-locking.

        Finally, the infinitesimal Isostable Response Curve (iIRC) explicitly addresses the strongly relaxing, return-to-rest dynamics of excitable nodes that possess a stable fixed point rather than a periodic orbit~\cite{wilsonExtendingPhaseReduction2015}. 
        While both the $\ePRC$ and the iIRC account for relaxation, the iIRC quantifies continuous-time trajectory shifts following an isolated state perturbation. 
        The $\ePRC$ abstracts away the continuous return trajectory entirely, providing a purely discrete input--output timing map for driven event propagation.

    \subsubsection{Excitatory vs. inhibitory ePRC}
    
        The $\ePRC$ profile differs qualitatively between excitatory and inhibitory perturbations (with nominal coupling remaining inhibitory throughout).
        Most importantly, excitatory inputs consistently produce phase \textit{advance} ($\phaseShift < 0$), whereas inhibitory coupling consistently induces phase \textit{delay} ($\phaseShift > 0$): by hyperpolarising the membrane and prolonging the time needed to reach threshold, it delays the subsequent output event. 
        This is a striking property of the event phase response curves in Figure~\ref{fig:eprc:example}.
        Note that this should not be interpreted as a mathematical property, but rather as a phenomenological consequence of the modelling assumptions: for the neuron and synapse model in this paper (see Appendix~\ref{sec:app:models}), and assuming the perturbation does not perturb the steady-state 1:1 transmission behaviour of the neuron.
        
        The stereotypical profile of the $\ePRC$ is shown for excitatory and inhibitory perturbations in Figures~\ref{fig:eprc:example}(a) and (b), along with a sample simulation showing the effect of the perturbation on the voltage trace in Figures~\ref{fig:eprc:example}(c) and (d). The clean complementarity between phase advance and phase delay enables bidirectional tuning of network phase advances and delays. This is a considerable simplification and novelty with respect to the classical framework of oscillator network analysis.

        We stress that specific neuronal models might exhibit regions of phase advance (delay) in their inhibitory (excitatory) ePRC, for specific choices of the parameters. 
        The classical literature often associates Type 1 PRCs with strictly positive phase advances~\cite{smealPhaseresponseCurvesSynchronized2010a, ermentroutTypeMembranesPhase1996, hanselSynchronyExcitatoryNeural1995}. 
        The strict absence of zero-crossings in both our excitatory and inhibitory ePRCs suggests that our underlying model exhibits Type 1 excitability~\cite{smealPhaseresponseCurvesSynchronized2010a}. 
        In any case, our main message is to stress the value of this property from a design perspective, rather than analysing whether this property may fail in specific parameter regimes. 
        With the models of this article, we observe this property as the rule rather than the exception, without fine-tuning of the parameters.

        \begin{figure}[htb]
  \centering
  \begin{tikzpicture}
    \pgfplotsset{
      every axis/.append style={
        grid=both,
        grid style={gray!15},
        minor tick num=0,
        tick align=inside,
        tick pos=left,
        yticklabel style={text width=2.5em, align=right},
        ylabel style={align=center, yshift=-10pt},
        label style={font=\small},
        tick label style={font=\footnotesize},
        title style={font=\small\bfseries},
        filter discard warning=false,
        unbounded coords=discard,
      }
    }

    \begin{groupplot}[
      group style={
        group size=2 by 2,
        horizontal sep=.8cm,
        vertical sep=1cm,
      },
      width=7cm,
      height=4.5cm,
      enlarge x limits=0.05,
    ]


      \nextgroupplot[
        title={Excitatory Perturbation},
        ylabel={Delay (ms)},
        xlabel={Phase offset $t_p$ (ms)},
      ]
      \addplot+[
        thick, 
        color=black, 
        mark=none
      ] table[x=tp, y=delta_t, col sep=comma] {plotdata/plotdata_eprc_exc_curve.csv};

      \nextgroupplot[
        title={Inhibitory Perturbation},
        xlabel={Phase offset $t_p$ (ms)},
      ]
      \addplot+[
        thick, 
        color=black, 
        mark=none
      ] table[x=tp, y=delta_t, col sep=comma] {plotdata/plotdata_eprc_inh_curve.csv};


      \nextgroupplot[
        ylabel={Voltage (mV)},
        xlabel={Time $t$ (ms)},
        xmin=230, xmax=280,
        legend style={
          font=\tiny,
          at={(0.98,0.98)},
          anchor=north east,
          draw=gray!40,
          fill=white,
          fill opacity=0.8,
          text opacity=1,
          legend image post style={scale=0.5}, 
          nodes={scale=0.9, transform shape}, 
          xstep=1pt,                           
        },
      ]
      \addplot[
        thick, 
        color=black
      ] table[x=time, y=v_nom, col sep=comma] {plotdata/plotdata_eprc_exc_trace_tp5.csv};
      \addlegendentry{$V_{\mathrm{nom}}$};

      \addplot[
        thick, 
        color=red!80!black
      ] table[x=time, y=v_pert, col sep=comma] {plotdata/plotdata_eprc_exc_trace_tp5.csv};
      \addlegendentry{$V_{\mathrm{p}}$};

      \addplot[
        semithick, 
        color=black, 
        dashed
      ] table[x=t_nom, y=y, col sep=comma] {plotdata/plotdata_eprc_events.csv};
      \addlegendentry{$\eventIn_{\mathrm{nom}}$};

      \addplot[
        semithick, 
        color=red!80!black, 
        dashed
      ] table[x=t_pert, y=y, col sep=comma] {plotdata/plotdata_eprc_events.csv};
      \addlegendentry{$\eventIn_{\mathrm{p}}$};

      \nextgroupplot[
        xlabel={Time $t$ (ms)},
        xmin=230, xmax=280,
        legend style={
          font=\tiny,
          at={(0.98,0.98)},
          anchor=north east,
          draw=gray!40,
          fill=white,
          fill opacity=0.8,
          text opacity=1,
          legend image post style={scale=0.5}, 
          nodes={scale=0.9, transform shape}, 
          xstep=1pt,                           
        },
      ]
      \addplot[
        thick, 
        color=black
      ] table[x=time, y=v_nom, col sep=comma] {plotdata/plotdata_eprc_inh_trace_tp5.csv};
      \addlegendentry{$V_{\mathrm{nom}}$};

      \addplot[
        thick, 
        color=red!80!black
      ] table[x=time, y=v_pert, col sep=comma] {plotdata/plotdata_eprc_inh_trace_tp5.csv};
      \addlegendentry{$V_{\mathrm{p}}$};

      \addplot[
        semithick, 
        color=black, 
        dashed
      ] table[x=t_nom, y=y, col sep=comma] {plotdata/plotdata_eprc_events.csv};
      \addlegendentry{$\eventIn_{\mathrm{nom}}$};

      \addplot[
        semithick, 
        color=red!80!black, 
        dashed
      ] table[x=t_pert, y=y, col sep=comma] {plotdata/plotdata_eprc_events.csv};
      \addlegendentry{$\eventIn_{\mathrm{p}}$};

    \end{groupplot}

    \node[anchor=south west, font=\small\bfseries\sffamily, inner sep=0pt] 
      at ([yshift=2pt, xshift=-15pt]group c1r1.north west) {(a)};
    \node[anchor=south west, font=\small\bfseries\sffamily, inner sep=0pt] 
      at ([yshift=2pt, xshift=-15pt]group c2r1.north west) {(b)};
    \node[anchor=south west, font=\small\bfseries\sffamily, inner sep=0pt] 
      at ([yshift=2pt, xshift=-15pt]group c1r2.north west) {(c)};
    \node[anchor=south west, font=\small\bfseries\sffamily, inner sep=0pt] 
      at ([yshift=2pt, xshift=-15pt]group c2r2.north west) {(d)};

  \end{tikzpicture}
  \caption{Event Phase Response Curves (ePRC) and corresponding voltage/pulse traces for excitatory (left, $\gSyn^{\mathrm{p,exc}}=0.05$) and inhibitory (right, $\gSyn^{\mathrm{p,inh}}=0.15$) perturbations. \textbf{(a)}, \textbf{(b)} Output timing shift $\Delta \timeOut{}$ as a function of the input phase offset $\phaseOffset$. \textbf{(c)}, \textbf{(d)} Nominal voltage baseline $V_{\mathrm{nom}}$ (solid black), perturbed trajectory $V_{\mathrm{p}}$ (solid red), and nominal $\eventIn_{\mathrm{nom}}$ (dashed black) and perturbing (dashed red) input events $\eventIn_{\mathrm{p}}$ for a phase offset $\phaseOffset=10$.}
  \label{fig:eprc:example}
\end{figure}

    \subsubsection{Phase resetting and entrainment dynamics}
    
        The $\ePRC$ governs the phase-resetting mechanism and dictates the stability of entrainment under external periodic forcing.
        The related phase transition map reveals critical phase transition points (e.g., borders of $1{:}1$ phase-locking, regions of event skip, or multi-spike regimes). 
        
        We can study simple entrainment ($\nominalT = \perturbT$) directly through the ePRC, following the standard approach of phase response theory: zero--crossings identify the fixed points of the phase transition map, while their slopes determine stability.
        From the ePRC obtained above, we can predict the state of entrainment of the node, given the configuration of the nominal and perturbing synapse and the respective intrinsic periods $\nominalT$ and $\perturbT$.
        However, since neither ePRC possesses a zero-crossing, stable-phase-locking is not predicted.
        On the other hand, phase-locking is achieved at fixed points $\phaseOffset^*$ satisfying $\ePRC(\phaseOffset^*) = \perturbT - \nominalT$ represents the period mismatch between the nominal oscillation and the perturbing signal.

\begin{figure}[htb]
  \centering
  \begin{tikzpicture}
    \definecolor{stimuluscolor}{RGB}{100, 100, 100} 

    \pgfplotsset{
      every axis/.append style={
        grid=both,
        grid style={gray!15},
        minor tick num=0,
        tick align=inside,
        tick pos=left,
        yticklabel style={text width=2.5em, align=right},
        ylabel style={align=center, yshift=-10pt},
        label style={font=\small},
        tick label style={font=\footnotesize},
        title style={font=\small\bfseries},
        filter discard warning=false,
        unbounded coords=discard,
      },
      neuron scatter/.style={
        scatter,
        scatter src=explicit symbolic,
        point meta=\thisrow{neuron_id},
        scatter/classes={
          0={node0color},
          1={node1color},
          2={node2color},
          3={node3color}
        }
      }
    }

    \begin{groupplot}[
      group style={
        group size=2 by 1,
        horizontal sep=1.2cm,
      },
      width=6.5cm,
      height=4.5cm,
      enlarge x limits=0.05,
    ]

      \nextgroupplot[
        xlabel={Spike time $t$ (ms)},
        ylabel={Neuron},
        ymin=-1, ymax=4,
        ytick={0,1,2,3},
        xmin=400, xmax=600,
        legend style={
          font=\tiny,
          at={(0.97, 0.97)},
          anchor=north east,
          draw=gray!40,
          fill=white,
          fill opacity=0.9,
          text opacity=1,
        },
      ]
      
      \addplot[
        mark=none,
        dashed, color=stimuluscolor, line width=0.5pt,
        forget plot,
        quiver={u=0, v=4, every arrow/.append style={-}} 
      ] table[
        x=last_sensory_time,
        y expr=-0.5,
        col sep=comma
      ] {plotdata/entrain_ring_spike_phase_data.csv};

      \addplot[
        only marks,
        mark=|,
        mark size=8pt,
        line width=0.8pt,
        neuron scatter,
        forget plot,
      ] table[
        x=spike_time,
        y=neuron_id,
        meta=neuron_id,
        col sep=comma
      ] {plotdata/entrain_ring_spike_phase_data.csv};

      \addlegendimage{dashed, color=stimuluscolor, line width=0.8pt}
      \addlegendentry{$E^\mathrm{P}$};

      \nextgroupplot[
        xlabel={Period Mismatch $\Delta T$ (ms)},
        ylabel={$\gSyn$},
        xmin=-10, xmax=4,
        ymin=0, ymax=0.10,
        ytick={0, 0.02, 0.04, 0.06, 0.08, 0.10},
        yticklabel style={/pgf/number format/fixed, /pgf/number format/precision=2},
        legend style={
          font=\tiny,
          at={(0.97, 0.97)},
          anchor=north east,
          draw=gray!40,
          fill=white,
          fill opacity=0.9,
          text opacity=1,
        },
      ]

      \addplot[
        fill=blue!20,
        opacity=0.6,
        draw=none,
        forget plot
      ] coordinates {
        (0.0, 0.000) (-0.5, 0.005) (-1.5, 0.010) (-2.0, 0.015) (-2.5, 0.020) (-3.0, 0.025) (-4.0, 0.030) (-4.0, 0.035) (-4.5, 0.040) (-5.0, 0.045) (-5.5, 0.050) (-5.5, 0.055) (-6.0, 0.060) (-6.0, 0.065) (-6.5, 0.070) (-6.5, 0.075) (-7.0, 0.080) (-7.0, 0.085) (-7.0, 0.090) (-5.0, 0.095)
        (-1.5, 0.095) (0.0, 0.090) (0.0, 0.085) (0.0, 0.080) (0.0, 0.075) (0.0, 0.070) (0.0, 0.065) (0.0, 0.060) (0.0, 0.055) (0.0, 0.050) (0.0, 0.045) (0.0, 0.040) (0.0, 0.035) (0.0, 0.030) (0.0, 0.025) (0.0, 0.020) (0.0, 0.015) (0.0, 0.010) (0.0, 0.005) (0.0, 0.000)
      } -- cycle;

      \addplot[
        thick,
        color=blue!80!black,
        forget plot
      ] coordinates {
        (0.0, 0.000) (-0.5, 0.005) (-1.5, 0.010) (-2.0, 0.015) (-2.5, 0.020) (-3.0, 0.025) (-4.0, 0.030) (-4.0, 0.035) (-4.5, 0.040) (-5.0, 0.045) (-5.5, 0.050) (-5.5, 0.055) (-6.0, 0.060) (-6.0, 0.065) (-6.5, 0.070) (-6.5, 0.075) (-7.0, 0.080) (-7.0, 0.085) (-7.0, 0.090) (-5.0, 0.095)
      };

      \addplot[
        thick,
        color=blue!80!black,
        forget plot
      ] coordinates {
        (-1.5, 0.095) (0.0, 0.090) (0.0, 0.085) (0.0, 0.080) (0.0, 0.075) (0.0, 0.070) (0.0, 0.065) (0.0, 0.060) (0.0, 0.055) (0.0, 0.050) (0.0, 0.045) (0.0, 0.040) (0.0, 0.035) (0.0, 0.030) (0.0, 0.025) (0.0, 0.020) (0.0, 0.015) (0.0, 0.010) (0.0, 0.005) (0.0, 0.000)
      };
      
      \addlegendimage{area legend, fill=blue!20, draw=blue!20}
      \addlegendentry{Entrainment};

      \addplot[
        only marks,
        mark=*,
        mark size=1pt,
        color=red,
      ] coordinates { (-4, 0.05) };
      \addlegendentry{Sample};

    \end{groupplot}

    \node[anchor=south west, font=\small\bfseries\sffamily, inner sep=0pt] 
      at ([yshift=2pt, xshift=-15pt]group c1r1.north west) {(a)};
    \node[anchor=south west, font=\small\bfseries\sffamily, inner sep=0pt] 
      at ([yshift=2pt, xshift=-15pt]group c2r1.north west) {(b)};

  \end{tikzpicture}
  \caption{\textbf{(a)} Spike raster plot of sample entrainment simulation showing neuronal spike times and sensory input events (dashed vertical lines, $\diffT=-4$ and $g_\mathrm{syn}=0.05$). \textbf{(b)} Entrainment region for excitatory perturbations indicating the sample simulation parameters (red dot).}
  \label{fig:eprc:entrain}
\end{figure}
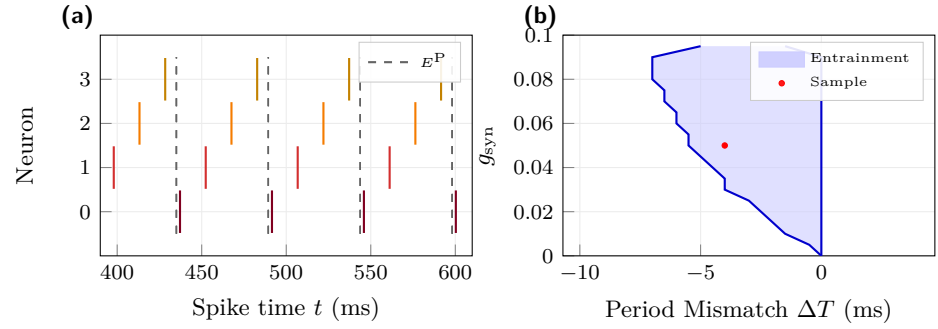

        Figure~\ref{fig:eprc:entrain}(a) shows the entraining effect of a perturbing input to the examined node ($\node{3}$) during network operation ($\netN = 4$) and a period mismatch $\diffT = \SI{-4}{\milli\second}$ for an intrinsic network period $\netT=60.9$. 
        The stable zero--crossing of the excitatory $\ePRC$, shifted by $\diffT$, is located at $\phaseOffset\approx 9$. As predicted, we observe stable phase-locking of the node to the perturbations, with a phase offset of $\phaseOffset^*=9.13$.
        
        This entrainment experiment can be repeated over a range of $\diffT$ and $\gSyn$ to find the boundaries of phase-locking. In Figure~\ref{fig:eprc:entrain}(b), the entrainment region of the network is shown for excitatory perturbations. 
        As expected, the entrainment region matches the maximum values of the ePRC ($\approx5.5$ for $\gSyn=0.05$).

    \subsection{Local tuning of the phase sensitivity}
        Analogous to the single-node event map analysis from the previous section, we examine how intrinsic parameter variations reshape nodal phase sensitivity.
        Because the maximum and minimum values of the $\ePRC$ directly determine the bounds of the entrainment region (and thus dictate the node's input sensitivity), modulating the amplitude of this curve provides direct control over the system's susceptibility to external forcing.

        \subsubsection{Single-parameter tuning}
            We first evaluate the impact of two key parameters on the amplitude and shape of the $\ePRC$.
            Just as in the previous section, we focus on $\tauDecay$ and $\gA$ for the sake of illustration.
    
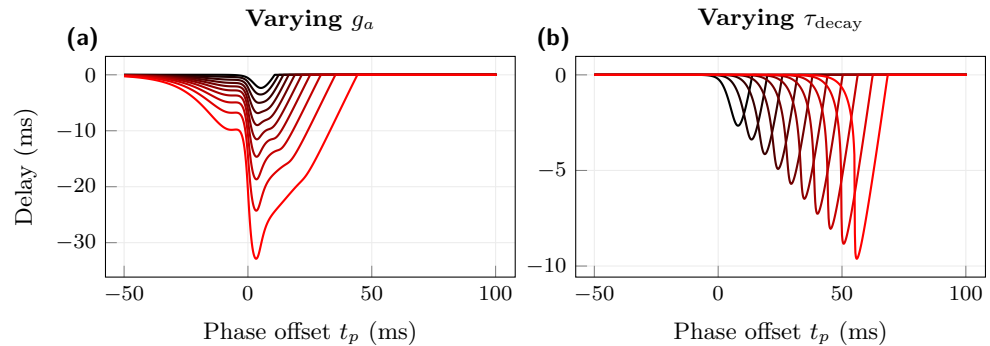
\begin{figure}[htb]
  \centering
  \begin{tikzpicture}
    \pgfplotsset{
      every axis/.append style={
        grid=both,
        grid style={gray!15},
        minor tick num=0,
        tick align=inside,
        tick pos=left,
        yticklabel style={text width=2.5em, align=right},
        ylabel style={align=center, yshift=-10pt},
        label style={font=\small},
        tick label style={font=\footnotesize},
        title style={font=\small\bfseries},
        filter discard warning=false,
        unbounded coords=discard,
      }
    }

    \begin{groupplot}[
      group style={
        group size=2 by 1,
        horizontal sep=.8cm,
      },
      width=7cm,
      height=4.5cm,
      enlarge x limits=0.05,
    ]

      \nextgroupplot[
        title={Varying $g_a$},
        ylabel={Delay (ms)},
        xlabel={Phase offset $t_p$ (ms)},
        legend style={
          font=\tiny,
          at={(0.97, 0.97)},
          anchor=north east,
          draw=gray!40,
          fill=white,
          fill opacity=0.9,
          text opacity=1,
          legend image post style={scale=0.5},
          nodes={scale=0.9, transform shape},
        },
      ]

      \foreach \i / \colpct in {0/0, 1/11, 2/22, 3/33, 4/44, 5/55, 6/66, 7/77, 8/88, 9/100} {
        \edef\temp{
          \noexpand\addplot[thick, color=red!\colpct!black, mark=none] 
            table[x=tp, y=val_\i, col sep=comma] {plotdata/plotdata_eprc_sweep_exc_neuron_g_a.csv};
        }
        \temp
      }

      \nextgroupplot[
        title={Varying $\tau_{\mathrm{decay}}$},
        xlabel={Phase offset $t_p$ (ms)},
        legend style={
          font=\tiny,
          at={(0.97, 0.97)},
          anchor=north east,
          draw=gray!40,
          fill=white,
          fill opacity=0.9,
          text opacity=1,
          legend image post style={scale=0.5},
          nodes={scale=0.9, transform shape},
        },
      ]

      \foreach \i / \colpct in {0/0, 1/11, 2/22, 3/33, 4/44, 5/55, 6/66, 7/77, 8/88, 9/100} {
        \edef\temp{
          \noexpand\addplot[thick, color=red!\colpct!black, mark=none] 
            table[x=tp, y=val_\i, col sep=comma] {plotdata/plotdata_eprc_sweep_exc_synapse_nom_tau_decay.csv};
        }
        \temp
      }

    \end{groupplot}

    \node[anchor=south west, font=\small\bfseries\sffamily, inner sep=0pt] 
      at ([yshift=2pt, xshift=-15pt]group c1r1.north west) {(a)};
    \node[anchor=south west, font=\small\bfseries\sffamily, inner sep=0pt] 
      at ([yshift=2pt, xshift=-15pt]group c2r1.north west) {(b)};

  \end{tikzpicture}
  \caption{Excitatory event phase response curves (ePRC) as a function of phase offset $t_p$ across varying intrinsic and synaptic parameters. \textbf{(a)} Effect of increasing $\iA$-conductance $g_a$ from $0.0$ (black) to $1.5$ (red) using an inverse exponential sweep. \textbf{(b)} Modulation across synaptic decay time constants $\tauDecay$ ranging from $3.0\,\mathrm{ms}$ (black) to $21.5\,\mathrm{ms}$ (red).}
  \label{fig:eprc:param_var}
\end{figure}

            Figure~\ref{fig:eprc:param_var} shows the excitatory $\ePRC$ across sweeps of $\gA$ (a) and $\tauDecay$ (b).
            Consistent with the nodal $\eDF$ analysis, both parameters directly govern the nominal onset delay, indicated by the rightmost boundary where the $\ePRC$ returns to zero. 
            However, modulation of the external (edge) parameter $\tauDecay$ of the nominal synapse quasi-linearly delays the nominal output (as established in Section~\ref{sec:endogenous}) while simultaneously deepening the $\ePRC$ and shifting its peak to later offsets. 
            In contrast, increasing the internal cellular parameter $\gA$ delays, deepens, and widens the $\ePRC$ in an exponential manner.
            Interestingly, its peak sensitivity remains centred at essentially the same phase offset, since the node becomes receptive as soon as the nominal synaptic current subsides.
            This qualitative distinction can be explained by the fact that the recovery of the neuron from inhibition is very sensitive during the phase of active $\iA$.
            As a consequence, small perturbations have a strong effect during this specific current activation.
            This is in direct contrast to the effect of $\tauDecay$, which effectively lengthens the time of full inhibition, during which the neuron is not sensitive at all.
    
        \subsubsection{Covaried tuning}
            Ideally, we wish to tune the sensitivity of a node independently of its endogenous behaviour by co-varying $\gA$ and $\tauDecay$.
            Because increasing $\gA$ enhances sensitivity at the cost of lengthening the nominal onset, concurrently decreasing $\tauDecay$ compensates for this shift, yielding an adjustable sensitivity level while maintaining identical nominal timing.
            
            In order to counteract the difference between the linear effect of $\tauDecay$ and the non-linear effect of $\gA$, we employ a Hill-type sigmoid modulation for $\gA$:
            \begin{equation}
                \gA(x) = \gA^+ \cdot \frac{x^a}{x^a + b \cdot (1 - x)^a}, \quad \text{with } b = \frac{\gA^+}{\gA^{\text{mid}}} - 1, \quad x \in [0, 1]
            \end{equation}
            where $\gA^-$ and $\gA^+$ are the lower and upper parameter bounds, $\gA^{\text{mid}}$ represents the midpoint value at $x = 0.5$, $b$ locks this midpoint constraint, and the exponent $a$ controls the steepness to balance the initial delay and late recovery.
            Here, the shaping parameters $a$ and $\gA^{\text{mid}}$ depend on the nodal dynamics and are fitted empirically.
            Concurrently, $\tauDecay$ decreases linearly along the sweep index $x$:
            \begin{equation}
                \tauDecay(x) = \tauDecay^+ - x \cdot (\tauDecay^+ - \tauDecay^-), \quad x \in [0, 1]
            \end{equation}
            where $\tauDecay^-$ and $\tauDecay^+$ are the lower and upper parameter values.
            
\begin{figure}[htb]
  \centering
  \begin{tikzpicture}
    \pgfplotsset{
      every axis/.append style={
        grid=both,
        grid style={gray!15},
        minor tick num=0,
        tick align=inside,
        tick pos=left,
        yticklabel style={text width=2.5em, align=right},
        ylabel style={align=center, yshift=-10pt},
        label style={font=\small},
        tick label style={font=\footnotesize},
        title style={font=\small\bfseries},
        filter discard warning=false,
        unbounded coords=discard,
      }
    }

    \begin{axis}[
      width=8.5cm,
      height=5.2cm,
      enlarge x limits=0.05,
      title={Covaried Sweep: $g_a$ vs. $\tau_{\mathrm{decay}}$},
      ylabel={Delay (ms)},
      xlabel={Phase offset $t_p$ (ms)},
      legend style={
        font=\tiny,
        at={(0.97, 0.03)},
        anchor=south east,
        draw=gray!40,
        fill=white,
        fill opacity=0.9,
        text opacity=1,
        legend image post style={scale=0.5},
        nodes={scale=0.9, transform shape},
      },
    ]

      \addplot[thick, color=black, mark=none] 
        table[x=tp, y=step1, col sep=comma] {plotdata/plotdata_eprc_covaried_exc_neuron_g_a_vs_synapse_nom_tau_decay.csv};
      \addlegendentry{$\gA=0.00, \tauDecay=20.50$};

      \addplot[thick, color=red!30!black, mark=none] 
        table[x=tp, y=step2, col sep=comma] {plotdata/plotdata_eprc_covaried_exc_neuron_g_a_vs_synapse_nom_tau_decay.csv};

      \addplot[thick, color=red!60!black, mark=none] 
        table[x=tp, y=step3, col sep=comma] {plotdata/plotdata_eprc_covaried_exc_neuron_g_a_vs_synapse_nom_tau_decay.csv};
      \addlegendentry{$\gA=1.04, \tauDecay=11.75$};

      \addplot[thick, color=red!85!black, mark=none] 
        table[x=tp, y=step4, col sep=comma] {plotdata/plotdata_eprc_covaried_exc_neuron_g_a_vs_synapse_nom_tau_decay.csv};

      \addplot[thick, color=red, mark=none] 
        table[x=tp, y=step5, col sep=comma] {plotdata/plotdata_eprc_covaried_exc_neuron_g_a_vs_synapse_nom_tau_decay.csv};
      \addlegendentry{$\gA=1.50, \tauDecay=3.00$};

    \end{axis}
  \end{tikzpicture}
  \caption{Event Phase Response Curves (ePRC) under opposite covariation of $\iA$-conductance $g_a$ and synaptic decay time constant $\tauDecay$. As $g_a$ increases according to a Hill-type sigmoid modulation from $0.00$ to $1.50$, $\tauDecay$ decreases linearly from $20.50\,\mathrm{ms}$ to $3.00\,\mathrm{ms}$, maintaining a constant nominal onset.}
  \label{fig:eprc:covary}
\end{figure}
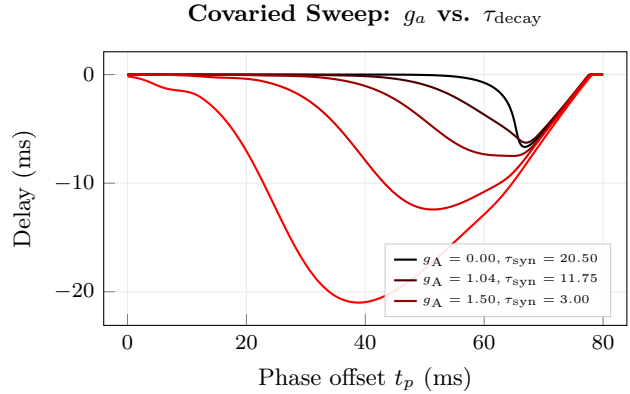

            This tuning, enabled by the A-type potassium current and synaptic decay time constant ($\iA$ and $\tauDecay$), is showcased in Figure~\ref{fig:eprc:covary}.
            As illustrated, sweeping the covariation index $x$ from 0 to 1 progressively deepens the phase advance, increasing the node's sensitivity to perturbations. 
            As desired, the rightmost boundary of the ePRC remains firmly anchored at a constant phase offset throughout the sweep. 
            This confirms that the covariation accurately preserves the nominal onset delay, successfully decoupling the node's exogenous sensitivity from its endogenous rhythmic timing.
            
    \subsection{Local tuning of the network interaction properties}
        
        We now extend our framework from single-node sensitivity to network-level exogenous regulation, using parameter heterogeneity as an explicit design tool.
        
        For the sake of illustration, we consider again the ring topology, and model interaction with the environment via additive synaptic inputs spread over all the neurons of the network. 
        Rebound neurons are mostly sensitive to external disturbances at the end of an inhibitory presynaptic event, meaning that only local perturbations can advance or delay the network rhythm. 
        In other words, the network sensitivity is not different from the neuron sensitivity.  
        
        However, coupling across the ring can introduce an additional timing cascade: when a perturbation shifts the output timing $\timeOut{i}$ of node $\node{i}$, the input arrival time $\timeIn{i+1}$ at the subsequent node $\node{i+1}$ is shifted by the identical amount.
        This consequently updates the effective perturbation phase offset $\phaseOffset^{i+1}$ for all downstream nodes in the sequence.
        
        Framed through the analogy of quadrupedal locomotion, where sensory feedback must selectively interface with specific limbs or gait phases, we examine three strategies for shaping this network sensitivity: (1) global homogeneous tuning, (2) single-node heterogeneous tuning, and (3) clustered functional tuning.
        
        \subsubsection{Homogeneous tuning}
            
            We first consider global homogeneous tuning, where all $N$ nodes receive identical sensory input (e.g., representing a shared feedback signal across all four limbs) and undergo identical parameter modulation. 

            We first illustrate how the entrainment of the network can be modulated without changing the nominal rhythm. 
        
\begin{figure}[htb]
  \centering
  \begin{tikzpicture}

    \begin{groupplot}[
      group style={
        group size=2 by 1,
        horizontal sep=1.0cm,
        y descriptions at=edge left,
      },
      width=7.2cm,
      height=3.8cm,
      unbounded coords=jump,
      tick label style={font=\footnotesize},
      label style={font=\small},
      title style={font=\small\bfseries},
      filter discard warning=false,
      scaled x ticks=false,
      every axis/.append style={
        grid=none,
        minor tick num=0,
        tick align=inside,
        tick pos=left, 
      },
    ]

      \nextgroupplot[
        title={Homogeneous},
        ylabel={Delay to stimulus (ms)},
        xlabel={Time (s)},
        xmin=0, xmax=10000,
        xtick={0,2500,5000,7500,10000},
        xticklabels={0,2.5,5,7.5,10},
        ymin=0, ymax=300,
        ytick={0,50,100,150,200,250,300},
        line width=0.5pt,
      ]
      \addplot+[color=node0color, mark=|, mark size=1.5pt, mark repeat=1, x filter/.expression={\thisrow{neuron_id} == 0 ? x : nan}] table[x=spike_time, y=delay_for_tikz, col sep=comma] {plotdata/entrain_ring_modulation_homo.csv};
      \addplot+[color=node1color, mark=|, mark size=1.5pt, mark repeat=1, x filter/.expression={\thisrow{neuron_id} == 1 ? x : nan}] table[x=spike_time, y=delay_for_tikz, col sep=comma] {plotdata/entrain_ring_modulation_homo.csv};
      \addplot+[color=node2color, mark=|, mark size=1.5pt, mark repeat=1, x filter/.expression={\thisrow{neuron_id} == 2 ? x : nan}] table[x=spike_time, y=delay_for_tikz, col sep=comma] {plotdata/entrain_ring_modulation_homo.csv};
      \addplot+[color=node3color, mark=|, mark size=1.5pt, mark repeat=1, x filter/.expression={\thisrow{neuron_id} == 3 ? x : nan}] table[x=spike_time, y=delay_for_tikz, col sep=comma] {plotdata/entrain_ring_modulation_homo.csv};
      
      \draw[gray, dashed, line width=2pt] (axis cs:2500,\pgfkeysvalueof{/pgfplots/ymin}) -- (axis cs:2500,\pgfkeysvalueof{/pgfplots/ymax});

      \nextgroupplot[
        title={Heterogeneous},
        xlabel={Time (s)},
        xmin=0, xmax=10000,
        xtick={0,2500,5000,7500,10000},
        xticklabels={0,2.5,5,7.5,10},
        ymin=0, ymax=300,
        ytick={0,50,100,150,200,250,300},
        line width=0.5pt,
      ]
      \addplot+[color=node0color, mark=|, mark size=1.5pt, mark repeat=1, x filter/.expression={\thisrow{neuron_id} == 0 ? x : nan}] table[x=spike_time, y=delay_for_tikz, col sep=comma] {plotdata/entrain_ring_modulation_hetero.csv};
      \addplot+[color=node1color, mark=|, mark size=1.5pt, mark repeat=1, x filter/.expression={\thisrow{neuron_id} == 1 ? x : nan}] table[x=spike_time, y=delay_for_tikz, col sep=comma] {plotdata/entrain_ring_modulation_hetero.csv};
      \addplot+[color=node2color, mark=|, mark size=1.5pt, mark repeat=1, x filter/.expression={\thisrow{neuron_id} == 2 ? x : nan}] table[x=spike_time, y=delay_for_tikz, col sep=comma] {plotdata/entrain_ring_modulation_hetero.csv};
      \addplot+[color=node3color, mark=|, mark size=1.5pt, mark repeat=1, x filter/.expression={\thisrow{neuron_id} == 3 ? x : nan}] table[x=spike_time, y=delay_for_tikz, col sep=comma] {plotdata/entrain_ring_modulation_hetero.csv};

      \draw[gray, dashed, line width=2pt] (axis cs:2500,\pgfkeysvalueof{/pgfplots/ymin}) -- (axis cs:2500,\pgfkeysvalueof{/pgfplots/ymax});

    \end{groupplot}

    \node[anchor=south west, font=\small\bfseries\sffamily, inner sep=0pt] 
      at ([yshift=5pt, xshift=-15pt]group c1r1.north west) {(a)};
    \node[anchor=south west, font=\small\bfseries\sffamily, inner sep=0pt] 
      at ([yshift=5pt, xshift=-15pt]group c2r1.north west) {(b)};

    \setlength{\tabcolsep}{3pt}
    \node[
      draw=black!80,
      fill=white,
      font=\footnotesize,
      anchor=north,
      inner sep=2pt
    ] at ($(group c1r1.south)!0.5!(group c2r1.south) - (0, 0.5cm)$) {
      \begin{tabular}{*{4}{c}}
        \tikz[baseline=-0.5ex]\draw[node0color, thick, mark=*, mark size=1.5pt] (0,0) -- (0.4,0); $n_0$ &
        \tikz[baseline=-0.5ex]\draw[node1color, thick, mark=square*, mark size=1.5pt] (0,0) -- (0.4,0); $n_1$ &
        \tikz[baseline=-0.5ex]\draw[node2color, thick, mark=triangle*, mark size=1.5pt] (0,0) -- (0.4,0); $n_2$ &
        \tikz[baseline=-0.5ex]\draw[node3color, thick, mark=diamond*, mark size=1.5pt] (0,0) -- (0.4,0); $n_3$
      \end{tabular}
    };

  \end{tikzpicture}
  \caption{Temporal evolution of nodal delays with respect to entraining inputs in a four-neuron ring network ($N = 4$). Vertical ticks indicate spike times. Panels contrast \textbf{(a)} homogeneous and \textbf{(b)} heterogeneous sensitivity tuning initiated at $t = 2500\text{ ms}$ through covariation of $\gA$ (from $0.0$ to $1.5$) and $\tauDecay$ (from $20.6$ to $3.0$). After homogeneous modulation \textbf{(a)}, all nodes $\node{i}$ are sensitive and can entrain to the entraining inputs. After heterogeneous modulation \textbf{(b)}, only node $\node{0}$ (dark red) becomes sensitive and is entrained.}
  \label{fig:eprc:tuning:ring}
\end{figure}
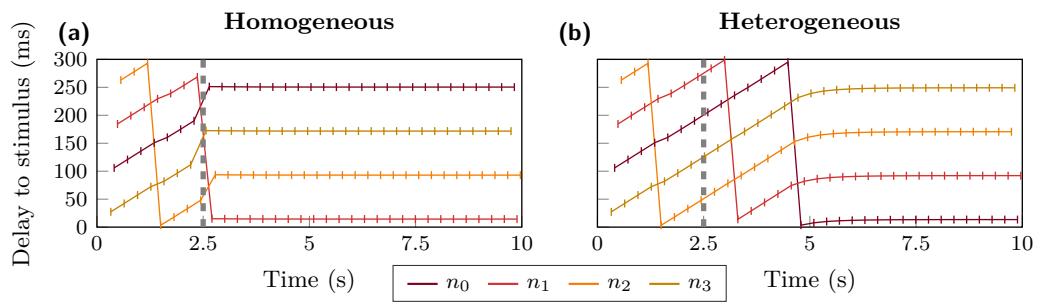

            Figure~\ref{fig:eprc:tuning:ring}(a) illustrates this mechanism: applying parameter covariation (increasing $\gA$ while concurrently decreasing $\tauDecay$) uniformly across all nodes enables the network to entrain to periodic inputs that fail to entrain the nominal, insensitive circuit.
            The uniform modulation scales the amplitude of every neuronal $\ePRC$ equally, increasing the sensitivity of the network across the entire ring.
            
            Because all nodes share identical sensitivity profiles, each position in the cycle is equally receptive to external driving. 
            This spatial symmetry creates $N$ multi-stable, phase-shifted entrainment equilibria, allowing the network to lock to incoming periodic signals at any of the $N$ phases.
        
        \subsubsection{Single-node heterogeneous tuning}    
        
            We next investigate how the modulation of a single node in an otherwise uniform network can modulate the network sensitivity.
            
            Following the analogy of quadrupedal locomotion, we consider phase-dependent sensory entrainment driven by periodic ground-contact events (such as a heel strike). 
            Rather than distributing sensitivity across all limbs, natural gaits often require sensory responsiveness to be restricted to a specific phase of the locomotor cycle. 
            Starting from a globally insensitive network, rendering node $\node{0}$ (representing the touch-down phase of a reference limb, such as the left forelimb) sensitive allows the entire network to phase-lock to the external sensory signal exclusively through that phase. 
            This single sensitive node acts as a primary phase-locking anchor, illustrating how parametric heterogeneity can dictate global entrainment. 
            This provides a computational analogue to phase-dependent reflex gating and sensory resetting in biological locomotion, where proprioceptive or tactile feedback entrains or resets the gait rhythm only when received within a narrow, functionally appropriate time window.

            While the nominal network completely ignores the periodic stimulus, Figure~\ref{fig:eprc:tuning:ring}(b) shows how modulating the sensitivity of a single node ($\node{0}$) enables entrainment of the entire network. 
            Upon making node $\node{0}$ sensitive, a single stable fixed point emerges in the network phase response exclusively at that phase. 
            Unlike the $N$ stable states produced by homogeneous tuning, this single-node modification selectively locks the network to a unique phase state, leaving all other phases insensitive to sensory perturbation.

        \subsubsection{Clustered and functional tuning}
        
            As a final illustration, we consider the modulation of a cluster of nodes to enable modular, multi-phase entrainment that maps directly to the clustered phase--task mapping established in Section~\ref{sec:endogenous}. 
            Grouping nodes into functional modules allows specific phases of an execution sequence to remain highly receptive to external feedback while keeping other phases autonomous and insensitive. 
            
            Returning to our quadrupedal locomotion analogy, the continuous muscular activity of a single limb can be partitioned into distinct sub-phases, each representing a different muscle activation profile during the stance phase. 
            By assigning a dedicated sequence of nodes to these sub-phases, their sensitivities to external inputs can be modulated entirely independently. 
            For instance, the initial impact sub-phase of the limb could be rendered insensitive to avoid disruption from transient mechanical noise upon touch-down, while a subsequent load-bearing sub-phase is tuned to be highly sensitive. 
            This functionally gated entrainment window allows incoming environmental cues to reliably phase-lock the rhythm only when the limb is dynamically ready to respond.
            
            We can also revisit the two-node functional clustering strategy introduced in Section~\ref{sec:3:clustering}. 
            While the basic framework in this work treats tasks as point events (or as remaining active until superseded by the next event), certain motor commands require explicit, fixed execution windows.
            By assigning dedicated nodes within a clustered phase, we can decouple phase transitions from phase durations.
            Specifically, using two nodes per limb phase allows a (sensitive) `transition node' to govern onset delay and entrainment bounds in response to external inputs (e.g., adapting touch-down timing to uneven terrain), while a downstream, (insensitive) `duration node' dictates the active execution time of the task (e.g., maintaining a fixed muscle contraction duration). 
            This maintains the simplicity of discrete event propagation rather than relying on continuous modulation of event durations.

            Much like the endogenous design problem, the clustered sensitivity strategies highlight the modular nature of the architecture. 
            They allow complex, multi-stage pattern generators to selectively interface with environmental signals only when sensory feedback is functionally relevant, without altering the underlying network connectivity.

\section{Conclusion}\label{sec:conclusion}

    In this work, we presented a distributed, event-based framework for rhythmic pattern generation in Central Pattern Generators (CPGs) composed of excitable, non-oscillatory neurons. 
    Our approach rests on a conceptual shift, adopting the perspective that biological rhythms can emerge from sequences of discrete events in excitable neuronal networks rather than from coupling between neuronal oscillators.  
    Our main goal was to illustrate that the the analysis and design of such rhythmic sequences is modular, robust, and scalable.  
    
    We addressed separately the \textit{internal} question of generating a particular rhythmic sequence in a network and the \textit{external} question of tuning how the internal rhythm can interact with the external environment. 
    The \textit{internal} question was analysed in Section \ref{sec:endogenous} by means of the event Describing Function ($\eDF$). 
    We showed how to modulate the global period and relative phases of a cyclic sequence through local control of the nodal transmission delays ($\nodeDelay{i}$). 
    The \textit{external} question was analysed in Section \ref{sec:exogenous} by means the event Phase Response Curve ($\ePRC$). 
    We showed how to modulate the entrainment properties of a cyclic network by local modulation of the nodal sensitivity. 
    The results illustrate that event-driven excitable architectures naturally accommodate heterogeneity, enabling flexible phase--task mappings, functional node clustering, and robust rhythm generation.

    All the illustrations in this paper were provided on the simplest example of rhythm, namely the generation of a periodic sequence in a ring network. 
    This restriction was adopted for the sake of illustration, but the proposed methodology is general and should be investigated in more general architectures. 
    In particular, two specific problems deserve further attention for the application of rhythm generation to rhythmic robotics.
    First, the analysis tools ($\eDF$ and $\ePRC$) should be generalised to bursting (rather than spiking) neurons. 
    Second, the generation of periodic sequences should be considered in network topologies that combine excitation and inhibition, such as the rebound winner-take-all networks studied in \cite{huoWinnerTakesAllMechanismEvent2025}. 
    The results in this paper provide encouraging first steps for the general question of generating hierarchical event sequences in neuromorphic control systems. 
    

%
%


\funding{The research leading to these results has received funding from the European Research Council under the Advanced ERC Grant Agreement SpikyControl n.101054323.}


\section*{Declaration of Generative AI in the Writing Process}
During the preparation of this work, the first author used Gemini (Google) to assist with language polishing and readability enhancements.

\data{The simulation code supporting this work are openly available on GitHub and permanently archived on Zenodo~\cite{wroblowskiBiophysicalSimulation2026} (https://doi.org/10.5281/zenodo.22939323).}

\begin{appendices}

\section{Mathematical models}\label{sec:app:models}

\subsection{Discrete-event node}
The discrete-event nodes framing our rhythmic network are implemented as a three-stage functional pipeline: an event-to-analog (E/A) synapse, a continuous-time conductance-based neuron, and an analog-to-event (A/E) detector (as illustrated in Figure~\ref{fig:ring_cut_expanded}). When discrete event timings arrive at the node, the E/A synapse converts them into a continuous synaptic current. This current drives the excitable neuron's state equations, generating a continuous membrane potential trajectory. Finally, the A/E event detector records a new discrete output event whenever this membrane potential crosses a predefined threshold. 

\subsection{Conductance-based mixed-feedback model}
The core excitable node used throughout the network simulations (figures 2--15) is a reduced two-dimensional conductance-based model governed by the mixed-feedback motif. The system state is defined by a fast membrane potential variable $V_f$ and a slower recovery variable $V_s$:
\begin{align}
    C_m \dot{V}_f &= -\iNa - I_{\mathrm{K}}  - I_{\mathrm{L}} - \iNaReb - \iA - I_{\mathrm{syn}} + I_{\mathrm{app}} \\
    \tau_s \dot{V}_s &= V_f - V_s
\end{align}
where $C_m$ is the membrane capacitance, $\tau_s$ is the slow time constant, $I_{\mathrm{app}}$ is the externally applied current, and $I_{\mathrm{syn}}$ is the sum of the incoming synaptic currents. 
The intrinsic ionic currents are formulated as:
\begin{align*}
    \iNa &= \gNa m_{\mathrm{Na}} h_{\mathrm{Na}} (V_f - E_{\mathrm{Na}}) \\
    I_{\mathrm{K}} &= \gK n_{\mathrm{K}} (V_f - E_{\mathrm{K}}) \\
    I_{\mathrm{L}} &= g_{\mathrm{L}} (V_f - E_{\mathrm{L}}) \\
    \iNaReb &= g_{\mathrm{Na,Reb}} m_{\mathrm{Na,Reb}} h_{\mathrm{Na,Reb}} (V_f - E_{\mathrm{Na,Reb}}) \\
    \iA &= \gA m_{\mathrm{A}} h_{\mathrm{A}} (V_f - E_{\mathrm{A}})
\end{align*}
Notice that we structurally decouple the transient sodium current into two distinct components: a high-threshold excitatory current ($\iNa$) and a low-threshold rebound current ($\iNaReb$). This allows for independent tuning of direct excitability and post-inhibitory rebound behaviour. 

For simplicity of tuning and neuromorphic design, the voltage-dependent activation ($m_x, n_x$) and inactivation ($h_x$) gating variables are defined using a generalised sigmoid function $S(x)$. 
While any standard sigmoidal curve would suffice, we explicitly define $S(x)$ as a saturating rectified linear unit (SatReLU) to provide strictly linear control boundaries:
\begin{equation}
    S(x) = \max(0, \min(1, x))
\end{equation}
The gating variables are then constructed using independent bias ($V_{\mathrm{bias}}$) and slope ($k$) parameters. 
Some gates ($m_{\mathrm{Na}}$, $m_{\mathrm{A}}$ and $m_{\mathrm{Na,Reb}}$) depend on the fast voltage $V_f$, while others ($h_{\mathrm{Na}}$, $h_{\mathrm{A}}$, $h_{\mathrm{Na,Reb}}$ and $n_{\mathrm{K}}$) depend on the slow recovery voltage $V_s$:
\begin{align}
    m_x(V_f) &= S\big(k_{m,x} (V_f - V_{m,x,\mathrm{bias}})\big) \\
    n_x(V_s) &= S\big(k_{n,x} (V_s - V_{n,x,\mathrm{bias}})\big) \\
    h_x(V_s) &= S\big(k_{h,x} (V_{h,x,\mathrm{bias}} - V_s)\big)
\end{align}

Unless otherwise specified in specific experiments (e.g., during parameter sweeps or structural modulations), the nominal baseline parameters for the node are set as follows. The membrane parameters are $C_m = 1.0 \, \mu\mathrm{F/cm^2}$ and $\tau_s = 5.0$ ms. The maximal conductances (in $\mathrm{mS/cm^2}$) are $\gNa = 50.0$, $g_{\mathrm{K}} = 1.0$, and $g_{\mathrm{L}} = 0.1$, with reversal potentials (in mV) of $E_{\mathrm{Na}} = 100.0$, $E_{\mathrm{K}} = -20.0$, and $E_{\mathrm{L}} = 0.0$. The A-type potassium current operates with $E_{\mathrm{A}} = -20.0$, while $\gA$ acts as a primary tuning variable (typically $\gA = 0.0$ for the baseline node, and $>1.0$ when specifically tuning transmission delays).

The gating kinetics (bias in mV, slope in $\mathrm{mV}^{-1}$) are:
\begin{itemize}
    \item \textbf{Na:} activation bias $10.0$, slope $0.02$; inactivation bias $30.0$, slope $0.05$.
    \item \textbf{Na,Reb:} activation bias $-10.0$, slope $0.1$; inactivation bias $-5.0$, slope $0.1$.
    \item \textbf{K:} activation bias $10.0$, slope $0.05$.
    \item \textbf{A:} activation bias $-15.0$, slope $0.1$; inactivation bias $-10.0$, slope $0.15$.
\end{itemize}
To transition the node from the excitable to the oscillatory regime, the rebound conductance is disabled ($g_{\mathrm{Na,Reb}} = 0.0$), the A-type conductance is elevated ($\gA = 5.076$) to retain consistent intrinsic network period for $\netN=4$, and a constant tonic drive of $I_{\mathrm{app}} = 4.0 \, \mu\mathrm{A/cm^2}$ is applied to bias the neuron into the oscillatory regime.

\subsection{Hodgkin-Huxley neuron model}
For the direct comparison of classical continuous dynamics versus event-driven dynamics in Figure~\ref{fig:closed_vs_open_neuron}, we employed a standard Hodgkin-Huxley model augmented with an A-type potassium current. The state equation is:
\begin{equation}
    C_m \dot{V} = I_{\mathrm{app}} - I_{\mathrm{Na}} - I_{\mathrm{K}} - I_{\mathrm{A}} - I_{\mathrm{L}} - I_{\mathrm{syn}}
\end{equation}
The gating variables evolve according to classical $\alpha/\beta$ exponential kinetics, which we shifted by $+65$ mV to align the nominal resting potential near $0$ mV. The maximal conductances are $G_{\mathrm{Na}} = 120.0$, $G_{\mathrm{K}} = 36.0$, $G_{\mathrm{A}} = 10.0$, and $G_{\mathrm{L}} = 0.3 \, \mathrm{mS/cm^2}$, with reversal potentials $E_{\mathrm{Na}} = 115.0$, $E_{\mathrm{K}} = -12.0$, and $E_{\mathrm{L}} = 10.6$ mV. The autonomous oscillatory regime is driven by a constant tonic current of $I_{\mathrm{app}} = 15.0 \, \mu\mathrm{A/cm^2}$, whereas the excitable regime receives $I_{\mathrm{app}} = 0.0$ alongside discrete synaptic event triggers.

\subsection{Event-based synapse}
Incoming discrete events trigger a dynamic synaptic conductance. To ensure smooth integration while retaining switch-like activation, the synaptic drive is governed by a logistic sigmoid based on the presynaptic membrane potential $V_{\mathrm{pre}}$:
\begin{equation}
    d(V_{\mathrm{pre}}) = \frac{1}{1 + \exp\left(-\frac{V_{\mathrm{pre}} - \vTh}{k_{\mathrm{syn}}}\right)}
\end{equation}
This drive activates the synaptic gating variable $s$ via a scaled rise and decay mechanism:
\begin{equation}
    \dot{s} = \alpha (1 - s) - \frac{1}{\tauDecay} s
\end{equation}
where $\alpha = \max\left(0, \left(\frac{1}{\tau_{\mathrm{rise}}} - \frac{1}{\tauDecay}\right) d(V_{\mathrm{pre}})\right)$. The resulting applied current on the postsynaptic neuron is $\iSyn = \gSyn s (V_f - E_{\mathrm{rev}})$. 

By default, network interactions rely on inhibitory coupling with $\gSyn = 3.0 \, \mathrm{mS/cm^2}$, $E_{\mathrm{rev}} = -20.0$ mV, $\vTh = 30.0$ mV, a steepness of $k_{\mathrm{syn}} = 0.1$, a fast rise time $\tau_{\mathrm{rise}} = 0.1$ ms, and a baseline decay time $\tauDecay = 2.15$ ms. Sensory perturbations (used in ePRC characterization) utilise a maximal synaptic conductance of $\gSyn = 0.05 \, \mathrm{mS/cm^2}$ with $E_{\mathrm{rev}} = 50.0$ mV for excitatory inputs, and $\gSyn = 0.15 \, \mathrm{mS/cm^2}$ with $E_{\mathrm{rev}} = -20.0$ mV for inhibitory inputs.

\subsection{Event detection}
The analog-to-event (A/E) conversion acts as the final stage of the nodal pipeline. A discrete output event is registered at time $t$ when the fast membrane potential $V_f(t)$ crosses a detection threshold $V_{\mathrm{detect}} = 30.0$ mV (or $50.0$ mV depending on the amplitude of the chosen parameter regime) with a strictly positive derivative. These discrete timestamps formulate the outgoing event sequence $\eventOut$ that propagates to downstream nodes.

\section{Simulation details}\label{sec:app:simulation}
All neuronal models and network architectures were simulated computationally using custom Python scripts. To accommodate the specific numerical requirements of different analyses, integration was performed using a Forward Euler scheme. For extended network simulations (figures 2, 3, 4, 8, 9, 10, 12, and 15), a time step of $dt = 0.01$ ms (or up to $0.025$ ms where stability permitted) provided sufficient resolution. However, evaluating the discrete-event mapping curves (the $\eDF$ and $\ePRC$ in figures 6, 7, 11, 13, and 14) required capturing minute, sub-millisecond phase shifts. For these characterizations, the integration step was reduced to $dt = 0.001$ or $0.0001$ ms. Furthermore, to strictly eliminate discretization artifacts when registering event timings, the exact moment of threshold crossing ($V_{\mathrm{detect}}$) was calculated between integration steps using linear interpolation. Simulation routines were accelerated via Just-In-Time (JIT) compilation using the \texttt{Numba} library.

\end{appendices}

\newpage

\bibliography{ref}

\end{document}